\documentclass[aps,superscriptaddress, twocolumn]{revtex4-2}
\usepackage{graphicx}% Include figure files
\usepackage{dcolumn}% Align table columns on decimal point
\usepackage{bm}% bold math
\usepackage{soul}  
\usepackage{xcolor} 
\usepackage{lineno}
\begin{document}

\title{Active turbulence in shear-thinning fluid}

\author{Hongyi Bian}
\affiliation{Global College, Shanghai Jiao Tong University, Shanghai 200240, People’s Republic of China}

\author{Chunhe Li}
\affiliation{Global College, Shanghai Jiao Tong University, Shanghai 200240, People’s Republic of China}

\author{Zixiang Lin}
\affiliation{Global College, Shanghai Jiao Tong University, Shanghai 200240, People’s Republic of China}

\author{Jin Zhu}
\affiliation{Global College, Shanghai Jiao Tong University, Shanghai 200240, People’s Republic of China}

\author{Weijie Chen}
\affiliation{Intelligent Medicine Institute, Shanghai Medical College, Fudan University, Shanghai 200032, People’s Republic of China}

\author{Gaojin Li}
\affiliation{State Key Laboratory of Ocean Engineering, School of Ocean \& Civil Engineering, Shanghai Jiao Tong
University, Shanghai 200240, People’s Republic of China}

\author{Yongxiang Huang}
\affiliation{State Key Laboratory of Marine Environmental Science \& Center for Marine Meteorology and Climate Change \&  College of Ocean and Earth Sciences, Xiamen University, Xiamen 361102, People’s Republic of China}

\author{Zijie Qu}
\thanks{Contact author: \href{mailto:zijie.qu@sjtu.edu.cn}{zijie.qu@sjtu.edu.cn}}
\affiliation{Global College, Shanghai Jiao Tong University, Shanghai 200240, People’s Republic of China}

\begin{abstract}
The study of active matter system has critical importance in revealing the physical essence of biological collective behavior. Dense bacterial suspension—a typical biological active matter, exhibits a wide range of phenomenona, among which bacterial turbulence has received extensive interest in recent years. This seemingly chaotic motion is widely studied in Newtonian fluid. However, studies based on complex fluids have predominantly focused on viscoelastic effects, leaving the role of shear-thinning viscosity largely unexplored despite its prevalence in natural bacterial environments like mucus and gastric fluids. Here, we experimentally employed Ficoll and Methocel polymers to study the impacts of various viscosities by Newtonian fluid and shear-thinning effects by non-Newtonian fluids on bacterial turbulence. We analyzed various physical properties, including energy, enstrophy, etc., and observed that the shear-thinning effect is significantly suppressed in highconcentration bacterial suspensions. While the ordered arrangement of polymer chains under shear flow leads to the microscopic anisotropic viscosity, the suppression is largely attributed to the disruption of polymer chains caused by strong interbacterial interactions in dense suspensions. To validate this hypothesis, we conducted experiments at a lower bacterial concentration and verified the findings using theoretical calculations based on the modified resistive force theory.
\end{abstract}

\maketitle

\section{Introduction}

Active matter systems are generally considered to be cluster systems produced by the individuals that can move autonomously, such as schooling fish \cite{lopez2012behavioural}, bird flocks \cite{cavagna2014bird}, and microtubule bundles with motor protein \cite{sanchez2012spontaneous}. One of the most distinctive features of active matter systems is their non-equilibrium state, sustained by continuous energy input and dissipation \cite{qi2022emergence}. The collective behavior of active matter systems is influenced by multiple factors, such as local interactions between individuals, external stimuli, and boundary conditions. Examples are emergent patterns \cite{liu2021density}, asters \cite{ross2019controlling}, vortices \cite{couzin2003self,lushi2014fluid,theillard2017geometric,liu2021viscoelastic}, and more. The active microbial system, including sperm \cite{creppy2015turbulence}, the artificial colloidal particle \cite{ackerman2017squirming,wu2021ion}, and the bacterial groups (such as \textit{Escherichia coli}), is an important branch of the study of active substances. The collective motion of the rod-like bacteria has been widely studied \cite{koch2011collective,be2019statistical,aranson2022bacterial,gachelin2014collective} and usually referred to as ``bacterial turbulence'' for its phenomenon similar to the inertial turbulence at high Reynolds number. 
% Studies of bacterial turbulence in Newtonian fluids have revealed many mechanisms of bacterial collective behavior. For example, the physical property, in-plane kinematic energy, was found to have a linear relationship with the enstrophy from the experiments \cite{dunkel2013fluid} carried out on the \textit{B. subtilis}. The energy spectrum was also found to have the universal scaling factor around -3 according to the experiments on the general active turbulence \cite{liu2021density,wensink2012meso,wang2017intrinsic,alert2022active}. Considering the density of bacteria, activity of bacteria, and the living-death ratio of bacteria as emergency conditions of the collective motion, experiments \cite {peng2021imaging} were carried out to yield a three-dimensional phase diagram, which provides verification data for current theories and quantitative data support for the follow-up experimental research. 

The study on the active system usually starts from the individual locomotion. Since microorganisms live in complex environments in nature, besides the Newtonian fluids, existing studies \cite {li2021microswimming} extends to the Non-Newtonian environments. Even the polymer solutions with constant viscosity in macroscopic tests can produce shear-thinning effects around high-shear bacterial flagella \cite{martinez2014flagellated}. Being the living environment of the widely studied bacterium \textit{E. coli}, gastric fluid exhibits both shear-thinning and viscoelastic properties \cite{pedersen2013characterization,ruiz2021overview} due to the presence of various polymers. Although the viscosity increases, an increase in bacterial swimming speed was often observed in non-Newtonian solutions. The unevenly distributed or anisotropic viscous stress exerted on bacteria \cite{magariyama2002mathematical,qu2020effects} caused by the shear-thinning effect, fewer bacterial tumbling events, and smaller wobble angle \cite{patteson2015running} caused by fluid elasticity are widely regarded as the main reasons for the increase in bacterial velocity. Fluid viscoelasticity can also prompt microorganisms to form collective motion by enhancing interactions among sperm cells, promoting the formation of clusters that facilitate successful fertilization \cite{tung2017fluid}.

Current research on bacterial turbulence in complex fluids remains limited, with most studies primarily focusing on viscoelastic effects. These investigations have revealed that viscoelastic solutions enhance mutual attraction and co-orientation among bacterial individuals \cite{li2016collective}, leading to enhanced temporal and spatial correlations \cite{ran2021polymers,liao2023viscoelasticity}. These structures, characterized by increased feature lengths, have been applied to control the periodic oscillating individual vortices within disk-shaped drops \cite{liu2021viscoelastic}. However, a gap exists in understanding the role of shear-thinning effects, which often coexist with viscoelasticity in fluids such as mucus, in bacterial turbulence. Although recent theoretical work predicts that shear-thinning effect alters the onset and transition of active turbulence, leading to complex phase behaviors \cite{reinken2025self}, its direct impact on the physical properties of an already developed turbulent state remains unclear. That limits a comprehensive understanding of bacterial dynamics in complex fluid environments.

In this study, we employed two polymer solutions—Ficoll and Methocel—to tune the rheological environment of the bacterial suspension. Ficoll consists of highly branched and compact molecules that do not form entangled or networked structures in solution, thus exhibiting a constant viscosity that is independent of shear rate. In contrast, Methocel is composed of linear or slightly branched polymer chains that can interact through hydrogen bonds \cite{yang2024molecular} or van der Waals forces, forming a weak transient network. When subjected to shear, these intermolecular interactions are progressively disrupted, leading to structural rearrangement within the polymer matrix and a corresponding decrease in viscosity, i.e., a shear-thinning behavior. Therefore, Methocel was selected to investigate the influence of shear-thinning on the collective bacterial dynamics, while Ficoll served as a Newtonian reference fluid due to its nearly constant viscosity \cite{qu2020effects}.

Both of the individual and collective motion of \textit{E. coli} in Newtonian and shear-thinning fluids are investigated, focusing on how the polymers and bacterial interactions influence turbulence dynamics. Several physical properties, including swimming speed, turbulence energy, energy flux, and characteristic (length and time) scales, are analyzed. Our findings reveal that the shear-thinning effect is significantly suppressed in high-concentration bacterial suspensions, leading to the non-monotonic energy change. To verify this observation, experiments are conducted at varying bacterial concentrations, and the change in turbulence energy is systematically compared. Furthermore, we adopt the modified Resistive Force Theory (RFT), which incorporates the influence of bacterial interaction on individual flagellar rotation, to calculate the energy changes in the suspension under different bacterial and polymer concentrations. This theoretical approach provides a quantitative framework to explain the observed suppression of shear-thinning effects in dense bacterial suspensions, offering new insights into the complex interplay between fluid rheology and collective bacterial behavior. 

\section{Results}

\begin{figure*}
    \centering
    \includegraphics[width=0.9\linewidth]{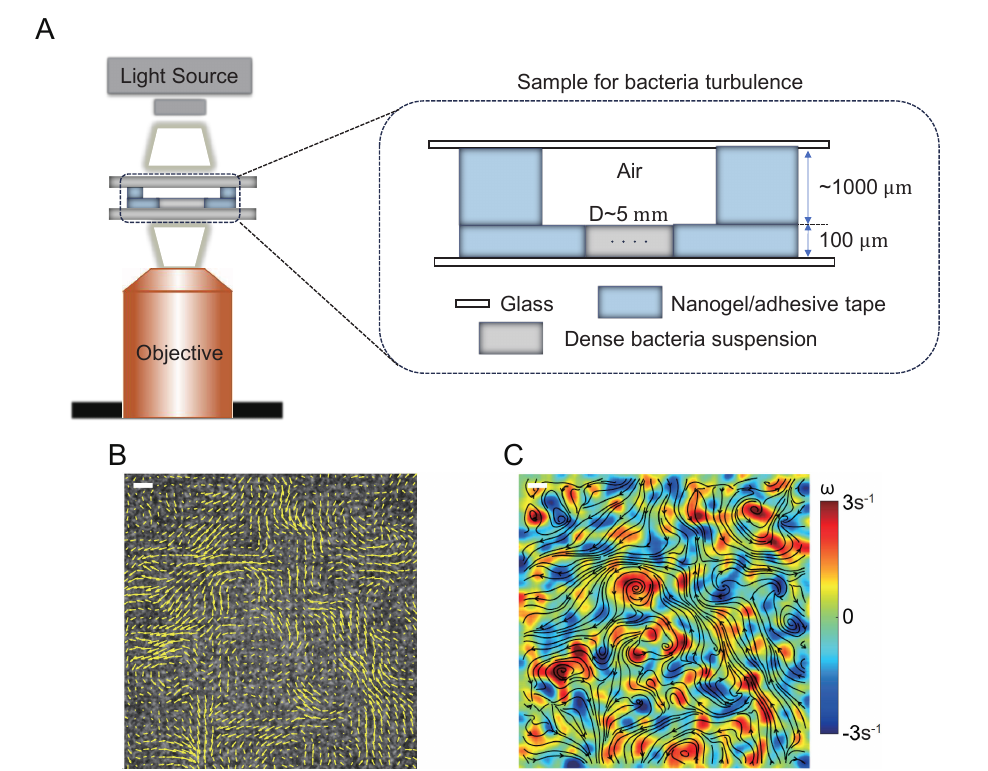}
    \caption{A) A brief schematic diagram of the experimental device and a detailed test fixture. The position of the objective is fixed throughout the experiments. B) Velocity flow field of the bacterial turbulence in pure buffer at a certain frame. The scale bar (white block) is $10 ~\mathrm{\mu m}$. C) Corresponding streamline and vorticity field of the velocity field in B).}
    \label{figure1}
\end{figure*}

Our experiments are carried out using PR+ \textit{E.coli} by the experimental setup shown in Fig.\ref{figure1} A (see Materials and Methods for more information) and videos are recorded at $60$ fps. The collective motion of the bacteria suspension, which shows local coherent flow and vortices (Fig. \ref{figure1} B and C) with scales much larger than a single bacteria, behaves similarly to the classical turbulent flows \cite{boffetta2012two,kraichnan1980two}. The velocity field of bacterial turbulence is calculated by the optical flow method \cite{cai2018dynamic}.

\begin{figure*}
    \centering
    \includegraphics[width=0.9\linewidth]{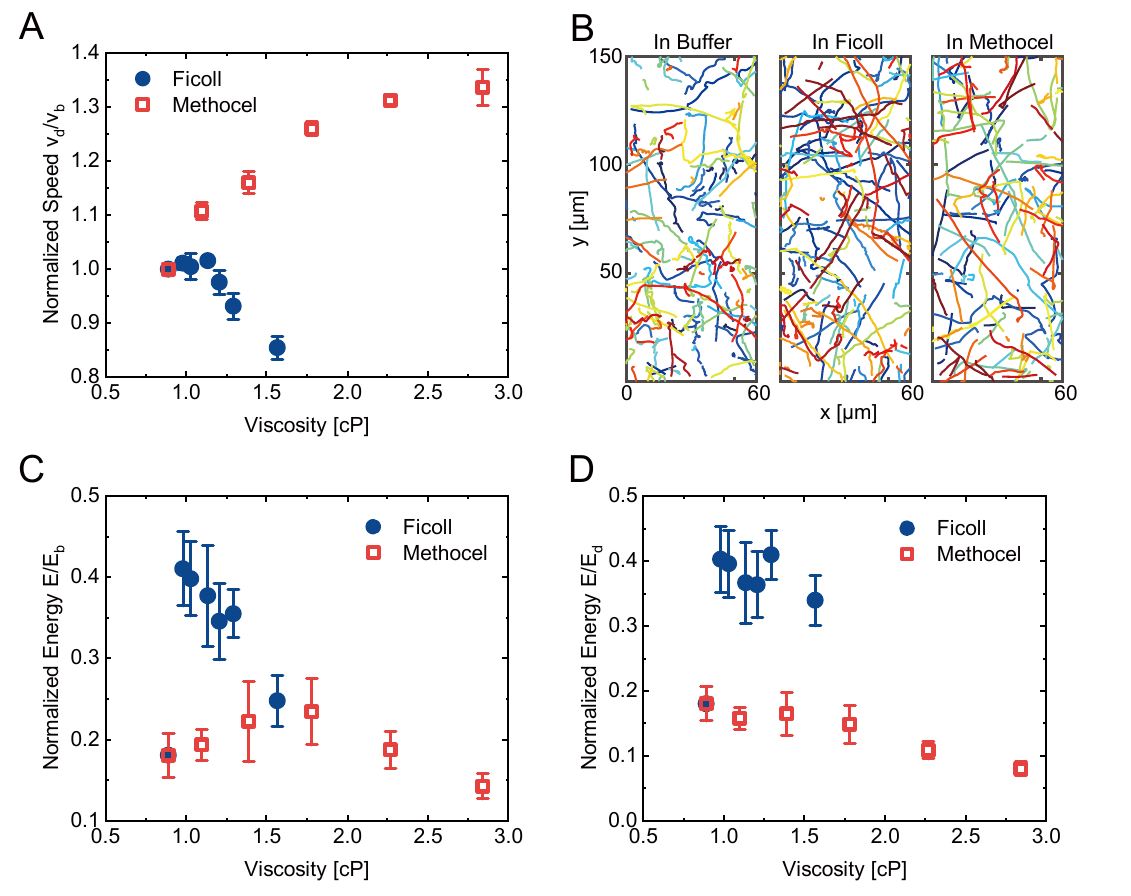}
    \caption{A) Normalized average speed of single swimmers in different polymer solutions. $\bar{v_b} = 18.1575 ~ \pm ~ 0.3364 ~\mathrm{\mu m /s}$. B) Trajectory of swimmers in Buffer, $7.5$\% Ficoll solution and $0.25$\% Methocel solution. The rougher trajectory of the bacteria is observed in Newtonian solution C) Averaged in-plane kinetic energy of bacterial turbulence normalized by the bacteria activity D) Energy of bacterial turbulence normalized by considering the bacteria averaged speed in polymer solution. All the error bars in this paper represent standard deviation.}
    \label{figure2}
\end{figure*}

Polymeric solutions are prepared by adding Ficoll and Methocel to the motility buffer. The Ficoll concentration ranges from 0 to $7.5$\% (wt/wt), which corresponds to a viscosity of $0.89$ to $1.5$ cP and the Methocel concentration is from $0$ to $0.25$\% (wt/wt) with a viscosity of $0.89$ to $3$ cP measured at a shear rate equal to $10 ~ \mathrm{s^{-1}}$ (see the rheological experimental data in Supplementary Information \cite{Bian2026Supplymental} Section 6). In response to the bacterial turbulence, the effect of polymers on individual bacteria swimming was also investigated experimentally. The swimming behavior of dilute \textit{E.coli} suspension was observed at a fixed ROI (see Materials and Methods for experimental equipment) and analyzed using ImageJ. Here, we use $v_d$ to indicate the average swimming speed of bacteria in dilute suspensions with different polymer concentrations and $v_b$ is the speed of bacteria in pure motility buffer. As shown in Fig. \ref{figure2} A, the addition of Ficoll initially enhances the speed of \textit{E.coli} and then decreases it. The normalized speed peaks at around $1.02$ while the lowest normalized speed reaches $0.85$. The initial growth is usually considered to be a result of the presence of small molecules in the polymer solution \cite{martinez2014flagellated,qu2020effects}, which may provide nutrition to bacteria. 

In the Methocel solutions, the bacteria's speed is increasing monotonically. From the swimming trajectory of bacteria shown in Fig. \ref{figure2} B, it is observed that the smaller swimming wobble angle leads to smoother paths. In Supplementary Information \cite{Bian2026Supplymental}, we quantify the smoothness of path by calculating the difference between the filtered trajectory and the original trajectory. Although the Methocel solution still behaves as a viscoelastic fluid, elasticity is not significant enough \cite{qu2020effects} to increase the speed of the bacteria by almost $40$ percent. Shear-thinning effect of the solution in conjunction with the high shear rate around the flagellar is the main reason for this phenomenon \cite{magariyama2002mathematical,qu2020effects}. 

The thickness of the dense bacterial suspension is around $100 ~\mathrm{\mu m}$, much larger than the individual bacterial cells, ensuring the formation of three-dimensional bacterial turbulence. We record the behavior of the collective motion near the bottom surface of the sample and calculate the in-plane kinetic energy (Fig. \ref{figure2} C) by $E = \langle (u^2+v^2)/2 \rangle$, where $u,v$ denote the velocity along $x,y$ direction, and also normalize it by considering the different bacteria colony activity $E_b = v_b^2/2$. It is found that the addition of a small amount of Ficoll to a bacterial suspension induces a significant abrupt change in bacterial turbulence activity (noticing that the Ficoll case and the Methocel case share the same point at viscosity equal to $1$ for the pure motility buffer). As the concentration of Ficoll further increases, the kinetic energy of the bacterial turbulence gradually decays. The initial abrupt change arises from the substantial reduction in bacterial activity during the centrifugation process used to concentrate the samples. The introduction of small Ficoll molecules can potentially provide energy and metabolites to damaged bacterial cells, facilitating their recovery and restoration of activity. The in-plane kinetic energy in shear-thinning Methocel solution does not exhibit a monotonic relationship with the polymer concentration, while the dilute bacteria solution shows a monotonic increasing speed. Instead, it reaches a peak at approximately $0.29$ and decreases subsequently. When the individual speed increases to $1.4$, the turbulence energy of the cluster drops to a value even smaller than that in a pure buffer.

To investigate the effect of individual movement speed on collective speed, energy is also non-dimensionalized with respect to the bacterial swimming speed in polymeric solution by $E_d = v_d^2/2$. It is observed that the ratio of turbulence energy and the energy $E_d$ approximately keeps unchanged ($\max{\delta (E/E_d)}\lesssim 15\%$) in both shear-thinning and Newtonian fluids at low viscosity (Fig. \ref{figure2} D), which indicates a linear correlation. However, the further addition of Methocel leads to a  significant decline ($\max{\delta (E/E_d)}\gtrsim 55\%$), which suggests the suppressed shear-thinning effect in high-density bacterial suspension. This explains the above results that although the shear-thinning effect causes an increase in the individual bacterial speed (Fig. \ref{figure2} A), the competition between the weakened shear-thinning effect and the increasing viscosity leads to a non-linear relationship (Fig. \ref{figure2} C) between turbulence energy and Methocel content.

To examine whether the shear rate caused by bacterial turbulence will lead to a significant decrease in viscosity, we calculate the in-plane shear rate $\varepsilon_{xy} = (dv/dx + du/dy)/2$. The magnitude of shear rate remains at the order of $0.1$ with the addition of polymers (Fig. \ref{figure3} A), and referring to the rheological properties of the Methocel solution measured (see Supplementary Information \cite{Bian2026Supplymental} Section 6), such shear rate is insufficient to induce a strong shear-thinning effect. Therefore, the impact of the turbulence shear scale on the polymer solution can be excluded. The enstrophy, $\Omega = \langle \omega^2 /2 \rangle$, which represents the rotation of the turbulence flow, has a trend similar to that of the shear rate (see Supplementary Information \cite{Bian2026Supplymental} Section 4). Here, $\omega = (dv/dx - du/dy)$ is vorticity.

\begin{figure*}
    \centering
    \includegraphics[width=0.9\linewidth]{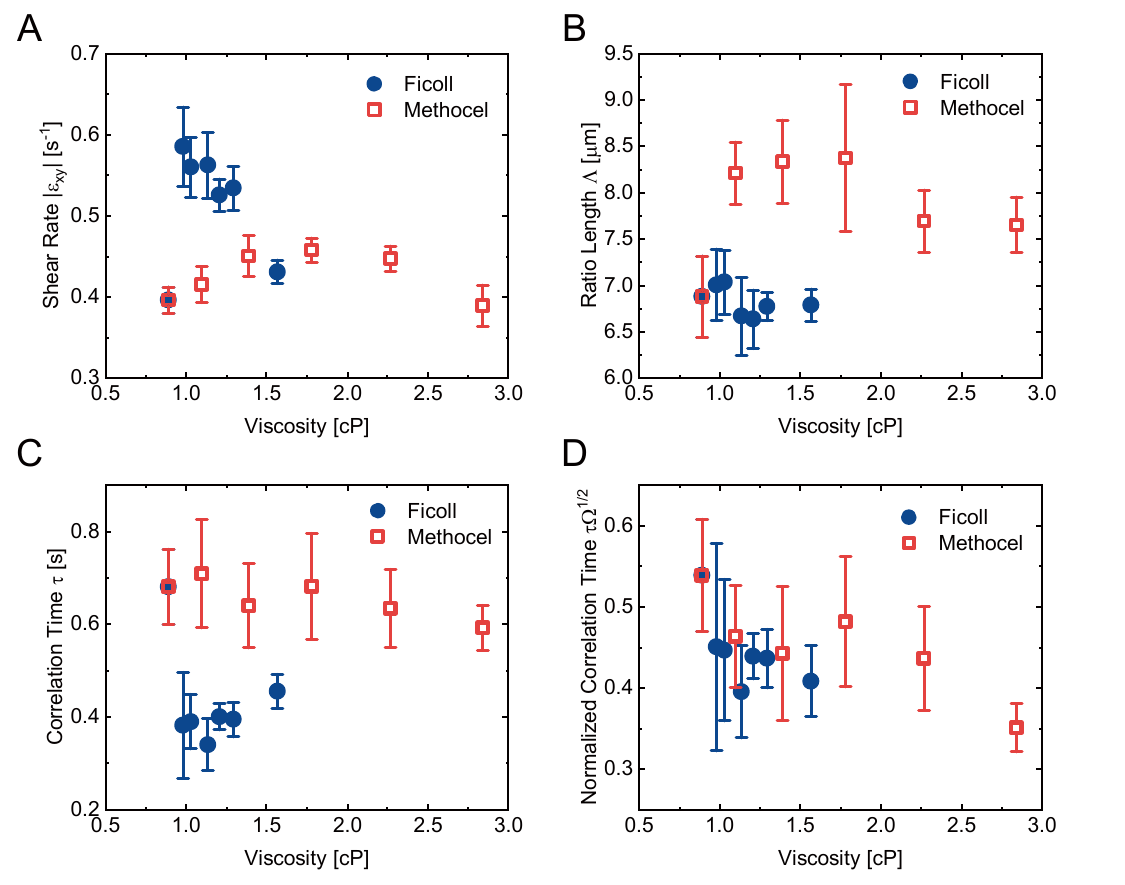}
    \caption{A) In-plane shear rate magnitude $|\varepsilon_{xy}|$ of bacterial turbulence in different polymer solutions. B), C) Ratio Length and Correlation time of bacterial turbulence in different polymer solutions. D) Normalized correlation time of bacterial turbulence in different polymer solutions. It is normalized by the energy and the turbulence structure size.}
    \label{figure3}
\end{figure*}

To further understand the properties of the bacterial turbulence, we calculate the enstrophy, which is highly related to the energy according to the previous experiments \cite{dunkel2013fluid, xie2022activity} on the bacterial turbulence with various bacterial activities. The ratio between the energy and enstrophy, shown as $\Lambda^2 = E/\Omega$, can represent the spatial structure of the bacterial turbulence, where we name $\Lambda$ the ratio length. With only viscosity changed, the ratio remains unchanged in the Ficoll solution (Fig. \ref{figure3} B) for our experiments. The slight memory effect from the viscoelasticity of the Methocel increases the characteristic length initially. However, the further addition of the Methocel does not keep increasing the characteristic size, which may result from the shear-thinning effect or from the projection of inherently three-dimensional structures onto a two-dimensional observation plane, suggesting the need for further three-dimensional reconstruction. We also report the correlation length (see Supplementary Information \cite{Bian2026Supplymental} Section 5), which is calculated from the velocity correlation function $C_u(r) = \langle \hat{v}(r_0)\cdot \hat{v}(r_0+r) \rangle_{r_0}$, where $\hat{v}$ is the normalized velocity field, and the trend is similar to the ratio length $\Lambda$ with magnitude two times larger. 

We examine the velocity correlation time using the temporal correlation function of the velocity, defined as $C_u(t) = \langle \hat{v}(t_0)\cdot \hat{v}(t_0+t) \rangle_{t_0}$. Then, the velocity correlation time $\tau$ is defined as the time when the correlation function $C_u(t)$ decays to $1/e$. The energy and correlation time in Ficoll solutions show opposite trends (Fig. \ref{figure2} C and Fig.\ref{figure3} C). The stronger turbulent flows (higher energy) speed up structural changes, as suggested in earlier studies \cite{ran2021polymers}. To analyze this, we calculate a dimensionless correlation time that combines effects of energy and structure size (Fig. \ref{figure3} D), which is $\tau E^{1/2} / \Lambda = \tau \Omega ^ {1/2}$, and this parameter decreases in both polymer solutions. While the tumbling rate would be regarded as rotational noise for bacterial turbulence close to threshold \cite{sokolov2012physical}, the reduction in normalized correlation time results from increased viscosity prolonging bacterial tumbling intervals.% However, in Methocel solutions, the initial normalized correlation time remains unchanged because of the persistent viscoelastic properties characteristic of Methocel. The memory effect of the fluid appears to suppress turbulent structure dissipation, consistent with previous experiments \cite{ran2021polymers}. When viscosity exceeds 1.75 cP, the parameter eventually decreases to values comparable with those observed in Ficoll solutions.

\begin{figure*}
    \centering
    \includegraphics[width=0.9\linewidth]{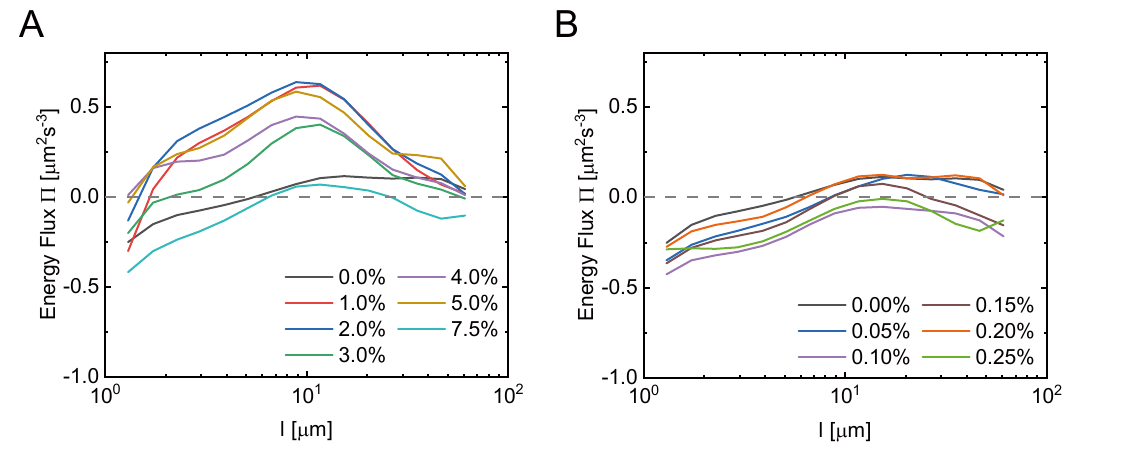}
    \caption{The energy flux of the bacterial turbulence in different polymer solutions. The negative value of the energy flux indicates the energy is transferred from the small scale to the large scale. A) Ficoll solution. B) Methocel solution.}
    \label{figure4}
\end{figure*}

The energy spectrum and the energy flux are calculated to further explore the energy distribution and the transfer in the bacterial turbulence. The energy spectrum is shown in the Supplementary Information \cite{Bian2026Supplymental} Section 7 and shows the scaling of $-0.8$ and $-2$ at different regions for the pure buffer. Then, the energy flux $\Pi(l) = - \sum_{i,j=1,2} \left[ (u_i u_j)^l - u_i^l u_j^l \right] \frac{\partial u_i^l}{\partial x_j}$, where the superscript $l$ denotes the original velocity field filtered by a low-pass filtering with size $l$ \cite{wang2017intrinsic}, is shown in Fig.\ref{figure4}. When no polymer is added, the energy flux shows a reasonable negative value since the energy injection of the system is from the bacteria flagella rotation at the micrometer scale. Then, when energy transfers from the small scale to the larger scale, it is dissipated by the viscous force and the flux is close to $0$. While the addition of the Methocel does not change the flux much, the Ficoll enhances its magnitude by increasing the overall bacterial turbulence energy. Except for the magnitude change, the intercept where the sign transitions from negative to positive shifts toward smaller scales. When we consider bacterial turbulence as a dry system, which ignores the solvent flow \cite{alert2022active}, there exists energy injection from bacteria, in-plane energy dissipation and the energy input or output between different bacteria layers (3D). Then, the energy injection from the higher layers would contribute to the positive energy flux since the velocity around the surface is restrained with the no-slip surface condition \cite{ishikawa2011energy}. According to the experiments (Fig.\ref{figure2} C), the energy of the system is much higher than the other when the concentration of Ficoll is from $1.0\%$ to $5.0\%$. The high velocity would lead to faster energy transfer, which should enhance the positive energy region. Also, while the structure of the turbulence is larger away from the surface \cite{ishikawa2011energy}, it is reasonable for the positive energy flux at a large scale. 

\section{Discussion}
Experiments reveal that the turbulence shear rate is insufficient to induce shear-thinning effects. Then, we study the shear-thinning mechanisms at a more microscopic level during bacterial collective motion. Previous studies \cite{qu2020effects} have shown that in polymer solutions, the rotation of individual bacterial flagella can generate strong local shear flows, reducing the viscosity of the fluid surrounding the flagella and significantly enhancing bacterial swimming speed in shear-thinning solutions. However, these studies focus primarily on dilute bacterial suspensions in which there are no significant hydrodynamic or physical interactions between individuals. This contrasts sharply with the characteristics of bacterial turbulence. Therefore, in bacterial turbulence, the strong interactions between individuals may weaken the local shear flows generated by the rotation of individual flagella or disrupt the regular alignment of polymer molecules (Methocel) in shear flows, thereby altering the shear-thinning effect.

\begin{figure*}
    \centering
    \includegraphics[width=0.9\linewidth]{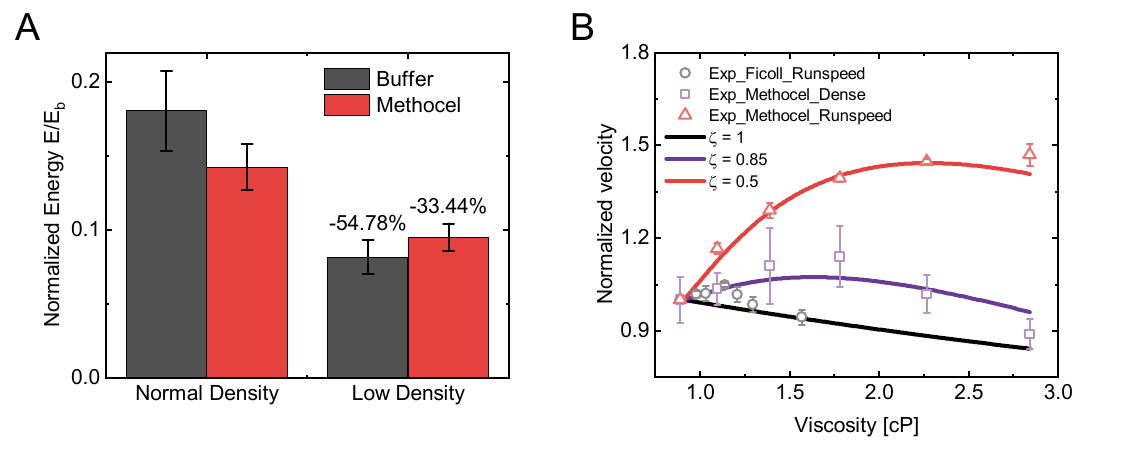}
    \caption{A) Normalized energy at different bacteria densities under buffer and $0.25$\% Methocel solution. B) Theoretical results of the bacteria swimming speed calculated by modified RFT. The light-colored points are the corresponding experimental results which are transferred to velocity units and normalized to start at $1$.}
    \label{figure5}
\end{figure*}

To verify this, we reduced the bacterial concentration in both pure buffer and high-concentration Methocel solutions and observed the changes in the turbulence energy. The experimental results are shown in Fig. \ref{figure5} A. At low concentration, the turbulence energy decreases in both solutions, but the magnitude of the reduction differs: in pure buffer, the energy decreases by $54.78$\%, while in the $0.25$\% Methocel solution, it decreases only by 33.44\%. This significantly smaller energy reduction (bootstrap test, one-sided, $p < 0.0001$) suggests that the bacterial concentration plays an important role in the shear-thinning effect at the microscopic scale: as the concentration decreases, the interactions between individuals weaken, making the shear-thinning effect more significant. In contrast, in high-concentration solutions, these interactions interrupt the formation of local shear flows, therefore limiting the shear-thinning effect in bacterial turbulence.

By introducing the modified resistive force theory (RFT) \cite{magariyama2002mathematical}, we can theoretically add the weakened shear-thinning effect induced by density change and reproduce the change in the injection rate of bacterial turbulent energy by individual swimming speed. The main idea here is to treat the viscosity as an anisotropic property. Because of the rearrangement of the polymers, the rotation of the cell body will not influence the polymer structure while the translational motion of the body would break the network and sustain a higher viscosity. So, two apparent viscosities are introduced as normal and tangential viscosity, $\mu_N$ and $\mu_T$. The tangential viscosity is generally smaller than the normal viscosity, which is the same case around the flagella. Here we assume the cell shear rate being 200 $s^{-1}$ and the flagellar shear rate around 4000 $s^{-1}$ \cite{qu2020effects}. Here, we also consider the force exerted on the flagella due to the slight viscoelasticity of the Methocel solution. The main calculation process is shown below \cite{magariyama2002mathematical} (for details, see Supplementary Information \cite{Bian2026Supplymental}). Applying the force- and torque-free condition of the flagella and cell body, we have

$$
\left[\begin{array}{ccc} 
    \alpha_c+\alpha_f &    0    & \gamma_f \\ 
    \gamma_f &    \beta_c   & \beta_f
\end{array}\right]
\left[\begin{array}{c} 
    v\\ 
    \omega_c\\ 
    \omega_f
\end{array}\right]
=
\left[\begin{array}{c} 
    0 \\ 
    0
\end{array}\right],
$$
where $\alpha,\beta,\gamma$ are the geometric constant related to viscosity and $v, \omega$ are velocity and angular speed with the subscript $f,c$ to distinguish flagella and cell body. Here, we also apply the constraint on the flagellum torque and its angular speed to complete the series of equations (see Supplementary Information \cite{Bian2026Supplymental} Section 8). Tangential viscosity is related to the bacteria density parameter $\zeta$:
$$
\mu_T = \mu_0+(\mu_N-\mu_0)*\zeta,
$$
where lower $\zeta$ indicates the stronger shear-thinning effect and $\mu_0 = 0.89~\mathrm{cP}$ is the viscosity of the water at $25^\circ \mathrm{C}$. The results of modified RFT in Fig. \ref{figure5} B are based on three different cases: (1) single bacteria swimming in the shear-thinning solution ($\zeta = 0.5$), (2) dense bacteria suspension in solution in which shear-thinning effect is restrained ($\zeta = 0.85$) and (3) single bacteria swimming in Newtonian solution ($\zeta = 1$). In both the first and third conditions, the results obtained by modified RFT are compared to the experimental results of characteristic run speed $U_0$, which have been transferred from the dilute bacteria swimming speed $\bar{U}$ according to the equation $U_0 = \bar{U} / (-0.185K+0.627)$ \cite{qu2018changes}, where $K$ is the speed distribution skewness. The modified RFT effectively reproduces the swimming speed trends of individual bacteria in shear-thinning and Newtonian fluids, despite the nutrition in Ficoll causes the derivation under viscosity of 1.25 $\mathrm{cP}$. In particular, during the transition phase between these two cases, the model captures the same non-monotonic variation observed in collective motion. Elevated values $\zeta$ indicate a weakening of the shear-thinning effect due to the increased concentration of the bacterial solution. The turbulence energy of the bacteria suspension in Methocel solution is squared to have the unit of micrometers. The experimental and theoretical results exhibit good agreement. 

\section{Conclusion}

In this work, we investigate the interplay between shear-thinning effects and bacterial turbulence in \textit{E. coli} suspensions using Newtonian (Ficoll) and shear-thinning (Methocel) polymer solutions. Key findings reveal that shear-thinning effects, which enhance individual bacterial swimming speed in dilute suspensions, are significantly suppressed in dense bacterial populations. Experimental results demonstrate a non-monotonic relationship between turbulence energy and polymer concentration in Methocel solutions, with turbulence energy peaking at intermediate concentrations before declining. This suppression is attributed to strong interactions in dense suspensions, which disrupt localized shear flows generated by flagellar motion and impede polymer rearrangement critical for shear-thinning behavior.

Without the strong elasticity, the characteristic length of the bacterial turbulence (e.g., the ratio between energy and enstrophy) changes slightly with the addition of polymers. The decreasing normalized correlation shows that the tumble of the bacteria enhanced by the higher viscosity could be regarded as an important noise in the bacterial turbulence near the turbulence threshold. From these results, we find the nonlinear relation between the turbulence structure size and normalized existence time, which is different from the experiments based on polymer solutions of quasi 2D bacterial turbulence \cite{ran2021polymers,liao2023viscoelasticity}. 

The modified Resistive Force Theory (RFT), incorporating anisotropic viscosity effects, successfully reproduces the observed trends of the in-plane energy, linking bacterial density to weakened shear-thinning responses. Experiments varying bacterial concentration further validate that reduced cell density amplifies shear-thinning effects, emphasizing the competition between collective interactions and fluid rheology.

While this work advances our understanding of bacterial dynamics in complex fluids, some limitations exist. The experimental protocol restricts bacterial concentrations near phase boundaries \cite{peng2021imaging}, necessitating future optimizations to explore broader density ranges. Additionally, the role of nanoscale polymer-bacterial interactions needs deeper investigation. The evolution of the structure of polymer solutions should be further explored to give valid proof to the assumption of anisotropic viscosity. These insights underscore the importance of integrating individual and collective behavior analyses to unravel microbial dynamics in physiologically relevant environments, paving the way for applications in biophysical control and synthetic active matter design.

\section{Materials and Methods}

\subsection{Cell Preparation}

The PR+ strain is the wild-type \textit{E. coli} (BW25113) engineered to express the light-driven proton pump proteorhodopsin (PR) \cite{liu2021density}. However, the light driven property is not activated during cultivation in this work and we treat it as normal wild type \textit{E. coli}. PR+ strain was initially streaked onto Lysogeny Broth (LB; 1 L water, 10 g tryptone, 5 g NaCl and 5 g yeast) medium agar plates from frozen stocks and was cultured for overnight (16 h) at 30 °C. Then a single colony of bacteria is selected and inoculated into 10 ml of Tryptone Broth (TB; 1 L of water, 10 g of tryptone and 5 g of NaCl) at 200 rpm for 16 h at 30 °C. To control the amount of bacteria, four 15 ml centrifuge tubes with 10 ml of TB were prepared. Then, we add 30 $\mu L$ bacterial suspension to each centrifuge tube to culture at 200 rpm for another 6 h. The suspension was then centrifuged and washed for three times (2500 rpm, 9 min for one time and 2200 rpm, 7 min for two times) with motility buffer (1 L of water, 11.2 g of K$_2$HPO$_4$, 4.8 g KH$_2$PO$_4$, 0.029 g EDTA, 3.9 g NaCl) to achieve the concentration of bacteria around 80 $n_0$ where $n_0 = 8\times 10^{8} ~\mathrm{mL}^{-1}$. Mix the suspension with the polymer solution (1:1) to get the final suspension. The number of experimental replicates for each condition is at least eight.

\subsection{Polymer Solutions}
To generate the solution with different rheological properties, we use two kinds of polymers: Ficoll 400 (Adamas, molecular weight 300k - 550k) and Methocel (Meryer, viscosity index 400$\mathrm{mPa\cdot s}$). The Ficoll powder and Methocel powder are mixed with the buffer at different ratios to achieve different concentrations. While the polymer solution is mixed with the bacteria suspension at the final step, the ratio of polymer cannot achieve the maximal solubility. The final ratio of the Ficoll solution is 0 to $7.5$\% (wt/wt) and the Methocel concentration is from $0$ to $0.25$\% (wt/wt). Rheology measurement was conducted using the TA Discovery HR20 multi-purpose rheometer with a 40.0 mm $1.0^\circ$ cone plate. The measured data is shown in Supplementary Information \cite{Bian2026Supplymental}. 

\subsection{Test Fixture}
The motion of cells in the dilute bacterial suspension was observed in a test fixture, which consisted of a 1 cm diameter well cut from a Deli nano tape of approximately 1 mm thickness. The sample was sandwiched between two coverslips. For the dense suspension, to supply oxygen and prevent evaporation, an additional 3M adhesive tape layer, 0.1 mm thick with a 5 mm diameter hole, was added below the nano tape layer used for the dilute sample. Approximately 2.5 $\mu L$ of dense suspension was then introduced into this small well. All chambers are fully sealed so that external conditions could not disrupt the inner flow fields.

\subsection{Imaging System and Data Analysis}
We used the Nikon Ti2-U microscope with a CFI S Plan Fluor ELWD 60X objective and a PCO Edge 5.5 sCMOS camera. The region of interest (ROI) was set to 1500 $\times$ 1500 pixels, allowing image collection at 60 fps. The corresponding image size was $162 \times 162 ~\mathrm{\mu m^2}$. To analyze the motility of individual bacteria, we used TrackMate \cite{ershov2022trackmate} in ImageJ to capture bacterial trajectories. For bacterial turbulence analysis, the flow field was generated using the Optical Flow method \cite{cai2018dynamic}. The main parameters used are as follows: $inter\_pass\_median\_filter = 5\times 5$ (median filtering in between pyramid levels), $pyramid\_level = 5$ (amount of coarse-to-fine steps), $iter = 2000$ (iteration steps), $\lambda_v = 1$ (smoothness parameter for velocity fields). MATLAB was then used for further analysis, including quantifying bacterial activity, turbulence energy, and other related metrics. 

\section{Acknowledgments}
We are grateful to Hepeng Zhang and Shuo Guo for their comments and discussions regarding this paper. We thank Shuo Guo and Xiang Cheng for kindly providing the bacterial strains used in this study. This research is funded by NSFC 12202275.


\begin{thebibliography}{49}%
\makeatletter
\providecommand \@ifxundefined [1]{%
 \@ifx{#1\undefined}
}%
\providecommand \@ifnum [1]{%
 \ifnum #1\expandafter \@firstoftwo
 \else \expandafter \@secondoftwo
 \fi
}%
\providecommand \@ifx [1]{%
 \ifx #1\expandafter \@firstoftwo
 \else \expandafter \@secondoftwo
 \fi
}%
\providecommand \natexlab [1]{#1}%
\providecommand \enquote  [1]{``#1''}%
\providecommand \bibnamefont  [1]{#1}%
\providecommand \bibfnamefont [1]{#1}%
\providecommand \citenamefont [1]{#1}%
\providecommand \href@noop [0]{\@secondoftwo}%
\providecommand \href [0]{\begingroup \@sanitize@url \@href}%
\providecommand \@href[1]{\@@startlink{#1}\@@href}%
\providecommand \@@href[1]{\endgroup#1\@@endlink}%
\providecommand \@sanitize@url [0]{\catcode `\\12\catcode `\$12\catcode `\&12\catcode `\#12\catcode `\^12\catcode `\_12\catcode `\%12\relax}%
\providecommand \@@startlink[1]{}%
\providecommand \@@endlink[0]{}%
\providecommand \url  [0]{\begingroup\@sanitize@url \@url }%
\providecommand \@url [1]{\endgroup\@href {#1}{\urlprefix }}%
\providecommand \urlprefix  [0]{URL }%
\providecommand \Eprint [0]{\href }%
\providecommand \doibase [0]{https://doi.org/}%
\providecommand \selectlanguage [0]{\@gobble}%
\providecommand \bibinfo  [0]{\@secondoftwo}%
\providecommand \bibfield  [0]{\@secondoftwo}%
\providecommand \translation [1]{[#1]}%
\providecommand \BibitemOpen [0]{}%
\providecommand \bibitemStop [0]{}%
\providecommand \bibitemNoStop [0]{.\EOS\space}%
\providecommand \EOS [0]{\spacefactor3000\relax}%
\providecommand \BibitemShut  [1]{\csname bibitem#1\endcsname}%
\let\auto@bib@innerbib\@empty
%</preamble>
\bibitem [{\citenamefont {Lopez}\ \emph {et~al.}(2012)\citenamefont {Lopez}, \citenamefont {Gautrais}, \citenamefont {Couzin},\ and\ \citenamefont {Theraulaz}}]{lopez2012behavioural}%
  \BibitemOpen
  \bibfield  {author} {\bibinfo {author} {\bibfnamefont {U.}~\bibnamefont {Lopez}}, \bibinfo {author} {\bibfnamefont {J.}~\bibnamefont {Gautrais}}, \bibinfo {author} {\bibfnamefont {I.~D.}\ \bibnamefont {Couzin}},\ and\ \bibinfo {author} {\bibfnamefont {G.}~\bibnamefont {Theraulaz}},\ }\bibfield  {title} {\bibinfo {title} {From behavioural analyses to models of collective motion in fish schools},\ }\href@noop {} {\bibfield  {journal} {\bibinfo  {journal} {Interface Focus}\ }\textbf {\bibinfo {volume} {2}},\ \bibinfo {pages} {693} (\bibinfo {year} {2012})}\BibitemShut {NoStop}%
\bibitem [{\citenamefont {Cavagna}\ and\ \citenamefont {Giardina}(2014)}]{cavagna2014bird}%
  \BibitemOpen
  \bibfield  {author} {\bibinfo {author} {\bibfnamefont {A.}~\bibnamefont {Cavagna}}\ and\ \bibinfo {author} {\bibfnamefont {I.}~\bibnamefont {Giardina}},\ }\bibfield  {title} {\bibinfo {title} {Bird flocks as condensed matter},\ }\href@noop {} {\bibfield  {journal} {\bibinfo  {journal} {Annu. Rev. Condens. Matter Phys.}\ }\textbf {\bibinfo {volume} {5}},\ \bibinfo {pages} {183} (\bibinfo {year} {2014})}\BibitemShut {NoStop}%
\bibitem [{\citenamefont {Sanchez}\ \emph {et~al.}(2012)\citenamefont {Sanchez}, \citenamefont {Chen}, \citenamefont {DeCamp}, \citenamefont {Heymann},\ and\ \citenamefont {Dogic}}]{sanchez2012spontaneous}%
  \BibitemOpen
  \bibfield  {author} {\bibinfo {author} {\bibfnamefont {T.}~\bibnamefont {Sanchez}}, \bibinfo {author} {\bibfnamefont {D.~T.}\ \bibnamefont {Chen}}, \bibinfo {author} {\bibfnamefont {S.~J.}\ \bibnamefont {DeCamp}}, \bibinfo {author} {\bibfnamefont {M.}~\bibnamefont {Heymann}},\ and\ \bibinfo {author} {\bibfnamefont {Z.}~\bibnamefont {Dogic}},\ }\bibfield  {title} {\bibinfo {title} {Spontaneous motion in hierarchically assembled active matter},\ }\href@noop {} {\bibfield  {journal} {\bibinfo  {journal} {Nature}\ }\textbf {\bibinfo {volume} {491}},\ \bibinfo {pages} {431} (\bibinfo {year} {2012})}\BibitemShut {NoStop}%
\bibitem [{\citenamefont {Qi}\ \emph {et~al.}(2022)\citenamefont {Qi}, \citenamefont {Westphal}, \citenamefont {Gompper},\ and\ \citenamefont {Winkler}}]{qi2022emergence}%
  \BibitemOpen
  \bibfield  {author} {\bibinfo {author} {\bibfnamefont {K.}~\bibnamefont {Qi}}, \bibinfo {author} {\bibfnamefont {E.}~\bibnamefont {Westphal}}, \bibinfo {author} {\bibfnamefont {G.}~\bibnamefont {Gompper}},\ and\ \bibinfo {author} {\bibfnamefont {R.~G.}\ \bibnamefont {Winkler}},\ }\bibfield  {title} {\bibinfo {title} {Emergence of active turbulence in microswimmer suspensions due to active hydrodynamic stress and volume exclusion},\ }\href@noop {} {\bibfield  {journal} {\bibinfo  {journal} {Communications Physics}\ }\textbf {\bibinfo {volume} {5}},\ \bibinfo {pages} {49} (\bibinfo {year} {2022})}\BibitemShut {NoStop}%
\bibitem [{\citenamefont {Liu}\ \emph {et~al.}(2021{\natexlab{a}})\citenamefont {Liu}, \citenamefont {Zeng}, \citenamefont {Ma},\ and\ \citenamefont {Cheng}}]{liu2021density}%
  \BibitemOpen
  \bibfield  {author} {\bibinfo {author} {\bibfnamefont {Z.}~\bibnamefont {Liu}}, \bibinfo {author} {\bibfnamefont {W.}~\bibnamefont {Zeng}}, \bibinfo {author} {\bibfnamefont {X.}~\bibnamefont {Ma}},\ and\ \bibinfo {author} {\bibfnamefont {X.}~\bibnamefont {Cheng}},\ }\bibfield  {title} {\bibinfo {title} {Density fluctuations and energy spectra of 3d bacterial suspensions},\ }\href@noop {} {\bibfield  {journal} {\bibinfo  {journal} {Soft Matter}\ }\textbf {\bibinfo {volume} {17}},\ \bibinfo {pages} {10806} (\bibinfo {year} {2021}{\natexlab{a}})}\BibitemShut {NoStop}%
\bibitem [{\citenamefont {Ross}\ \emph {et~al.}(2019)\citenamefont {Ross}, \citenamefont {Lee}, \citenamefont {Qu}, \citenamefont {Banks}, \citenamefont {Phillips},\ and\ \citenamefont {Thomson}}]{ross2019controlling}%
  \BibitemOpen
  \bibfield  {author} {\bibinfo {author} {\bibfnamefont {T.~D.}\ \bibnamefont {Ross}}, \bibinfo {author} {\bibfnamefont {H.~J.}\ \bibnamefont {Lee}}, \bibinfo {author} {\bibfnamefont {Z.}~\bibnamefont {Qu}}, \bibinfo {author} {\bibfnamefont {R.~A.}\ \bibnamefont {Banks}}, \bibinfo {author} {\bibfnamefont {R.}~\bibnamefont {Phillips}},\ and\ \bibinfo {author} {\bibfnamefont {M.}~\bibnamefont {Thomson}},\ }\bibfield  {title} {\bibinfo {title} {Controlling organization and forces in active matter through optically defined boundaries},\ }\href@noop {} {\bibfield  {journal} {\bibinfo  {journal} {Nature}\ }\textbf {\bibinfo {volume} {572}},\ \bibinfo {pages} {224} (\bibinfo {year} {2019})}\BibitemShut {NoStop}%
\bibitem [{\citenamefont {Couzin}\ and\ \citenamefont {Franks}(2003)}]{couzin2003self}%
  \BibitemOpen
  \bibfield  {author} {\bibinfo {author} {\bibfnamefont {I.~D.}\ \bibnamefont {Couzin}}\ and\ \bibinfo {author} {\bibfnamefont {N.~R.}\ \bibnamefont {Franks}},\ }\bibfield  {title} {\bibinfo {title} {Self-organized lane formation and optimized traffic flow in army ants},\ }\href@noop {} {\bibfield  {journal} {\bibinfo  {journal} {Proceedings of the Royal Society of London. Series B: Biological Sciences}\ }\textbf {\bibinfo {volume} {270}},\ \bibinfo {pages} {139} (\bibinfo {year} {2003})}\BibitemShut {NoStop}%
\bibitem [{\citenamefont {Lushi}\ \emph {et~al.}(2014)\citenamefont {Lushi}, \citenamefont {Wioland},\ and\ \citenamefont {Goldstein}}]{lushi2014fluid}%
  \BibitemOpen
  \bibfield  {author} {\bibinfo {author} {\bibfnamefont {E.}~\bibnamefont {Lushi}}, \bibinfo {author} {\bibfnamefont {H.}~\bibnamefont {Wioland}},\ and\ \bibinfo {author} {\bibfnamefont {R.~E.}\ \bibnamefont {Goldstein}},\ }\bibfield  {title} {\bibinfo {title} {Fluid flows created by swimming bacteria drive self-organization in confined suspensions},\ }\href@noop {} {\bibfield  {journal} {\bibinfo  {journal} {Proceedings of the National Academy of Sciences}\ }\textbf {\bibinfo {volume} {111}},\ \bibinfo {pages} {9733} (\bibinfo {year} {2014})}\BibitemShut {NoStop}%
\bibitem [{\citenamefont {Theillard}\ \emph {et~al.}(2017)\citenamefont {Theillard}, \citenamefont {Alonso-Matilla},\ and\ \citenamefont {Saintillan}}]{theillard2017geometric}%
  \BibitemOpen
  \bibfield  {author} {\bibinfo {author} {\bibfnamefont {M.}~\bibnamefont {Theillard}}, \bibinfo {author} {\bibfnamefont {R.}~\bibnamefont {Alonso-Matilla}},\ and\ \bibinfo {author} {\bibfnamefont {D.}~\bibnamefont {Saintillan}},\ }\bibfield  {title} {\bibinfo {title} {Geometric control of active collective motion},\ }\href@noop {} {\bibfield  {journal} {\bibinfo  {journal} {Soft Matter}\ }\textbf {\bibinfo {volume} {13}},\ \bibinfo {pages} {363} (\bibinfo {year} {2017})}\BibitemShut {NoStop}%
\bibitem [{\citenamefont {Liu}\ \emph {et~al.}(2021{\natexlab{b}})\citenamefont {Liu}, \citenamefont {Shankar}, \citenamefont {Marchetti},\ and\ \citenamefont {Wu}}]{liu2021viscoelastic}%
  \BibitemOpen
  \bibfield  {author} {\bibinfo {author} {\bibfnamefont {S.}~\bibnamefont {Liu}}, \bibinfo {author} {\bibfnamefont {S.}~\bibnamefont {Shankar}}, \bibinfo {author} {\bibfnamefont {M.~C.}\ \bibnamefont {Marchetti}},\ and\ \bibinfo {author} {\bibfnamefont {Y.}~\bibnamefont {Wu}},\ }\bibfield  {title} {\bibinfo {title} {Viscoelastic control of spatiotemporal order in bacterial active matter},\ }\href@noop {} {\bibfield  {journal} {\bibinfo  {journal} {Nature}\ }\textbf {\bibinfo {volume} {590}},\ \bibinfo {pages} {80} (\bibinfo {year} {2021}{\natexlab{b}})}\BibitemShut {NoStop}%
\bibitem [{\citenamefont {Creppy}\ \emph {et~al.}(2015)\citenamefont {Creppy}, \citenamefont {Praud}, \citenamefont {Druart}, \citenamefont {Kohnke},\ and\ \citenamefont {Plourabou{\'e}}}]{creppy2015turbulence}%
  \BibitemOpen
  \bibfield  {author} {\bibinfo {author} {\bibfnamefont {A.}~\bibnamefont {Creppy}}, \bibinfo {author} {\bibfnamefont {O.}~\bibnamefont {Praud}}, \bibinfo {author} {\bibfnamefont {X.}~\bibnamefont {Druart}}, \bibinfo {author} {\bibfnamefont {P.~L.}\ \bibnamefont {Kohnke}},\ and\ \bibinfo {author} {\bibfnamefont {F.}~\bibnamefont {Plourabou{\'e}}},\ }\bibfield  {title} {\bibinfo {title} {Turbulence of swarming sperm},\ }\href@noop {} {\bibfield  {journal} {\bibinfo  {journal} {Physical Review E}\ }\textbf {\bibinfo {volume} {92}},\ \bibinfo {pages} {032722} (\bibinfo {year} {2015})}\BibitemShut {NoStop}%
\bibitem [{\citenamefont {Ackerman}\ \emph {et~al.}(2017)\citenamefont {Ackerman}, \citenamefont {Boyle},\ and\ \citenamefont {Smalyukh}}]{ackerman2017squirming}%
  \BibitemOpen
  \bibfield  {author} {\bibinfo {author} {\bibfnamefont {P.~J.}\ \bibnamefont {Ackerman}}, \bibinfo {author} {\bibfnamefont {T.}~\bibnamefont {Boyle}},\ and\ \bibinfo {author} {\bibfnamefont {I.~I.}\ \bibnamefont {Smalyukh}},\ }\bibfield  {title} {\bibinfo {title} {Squirming motion of baby skyrmions in nematic fluids},\ }\href@noop {} {\bibfield  {journal} {\bibinfo  {journal} {Nature Communications}\ }\textbf {\bibinfo {volume} {8}},\ \bibinfo {pages} {673} (\bibinfo {year} {2017})}\BibitemShut {NoStop}%
\bibitem [{\citenamefont {Wu}\ \emph {et~al.}(2021)\citenamefont {Wu}, \citenamefont {Dai}, \citenamefont {Li}, \citenamefont {Gao}, \citenamefont {Wang}, \citenamefont {Liu}, \citenamefont {Zheng}, \citenamefont {Zhan}, \citenamefont {Chen}, \citenamefont {Cheng} \emph {et~al.}}]{wu2021ion}%
  \BibitemOpen
  \bibfield  {author} {\bibinfo {author} {\bibfnamefont {C.}~\bibnamefont {Wu}}, \bibinfo {author} {\bibfnamefont {J.}~\bibnamefont {Dai}}, \bibinfo {author} {\bibfnamefont {X.}~\bibnamefont {Li}}, \bibinfo {author} {\bibfnamefont {L.}~\bibnamefont {Gao}}, \bibinfo {author} {\bibfnamefont {J.}~\bibnamefont {Wang}}, \bibinfo {author} {\bibfnamefont {J.}~\bibnamefont {Liu}}, \bibinfo {author} {\bibfnamefont {J.}~\bibnamefont {Zheng}}, \bibinfo {author} {\bibfnamefont {X.}~\bibnamefont {Zhan}}, \bibinfo {author} {\bibfnamefont {J.}~\bibnamefont {Chen}}, \bibinfo {author} {\bibfnamefont {X.}~\bibnamefont {Cheng}}, \emph {et~al.},\ }\bibfield  {title} {\bibinfo {title} {Ion-exchange enabled synthetic swarm},\ }\href@noop {} {\bibfield  {journal} {\bibinfo  {journal} {Nature Nanotechnology}\ }\textbf {\bibinfo {volume} {16}},\ \bibinfo {pages} {288} (\bibinfo {year} {2021})}\BibitemShut {NoStop}%
\bibitem [{\citenamefont {Koch}\ and\ \citenamefont {Subramanian}(2011)}]{koch2011collective}%
  \BibitemOpen
  \bibfield  {author} {\bibinfo {author} {\bibfnamefont {D.~L.}\ \bibnamefont {Koch}}\ and\ \bibinfo {author} {\bibfnamefont {G.}~\bibnamefont {Subramanian}},\ }\bibfield  {title} {\bibinfo {title} {Collective hydrodynamics of swimming microorganisms: living fluids},\ }\href@noop {} {\bibfield  {journal} {\bibinfo  {journal} {Annual Review of Fluid Mechanics}\ }\textbf {\bibinfo {volume} {43}},\ \bibinfo {pages} {637} (\bibinfo {year} {2011})}\BibitemShut {NoStop}%
\bibitem [{\citenamefont {Be’er}\ and\ \citenamefont {Ariel}(2019)}]{be2019statistical}%
  \BibitemOpen
  \bibfield  {author} {\bibinfo {author} {\bibfnamefont {A.}~\bibnamefont {Be’er}}\ and\ \bibinfo {author} {\bibfnamefont {G.}~\bibnamefont {Ariel}},\ }\bibfield  {title} {\bibinfo {title} {A statistical physics view of swarming bacteria},\ }\href@noop {} {\bibfield  {journal} {\bibinfo  {journal} {Movement Ecology}\ }\textbf {\bibinfo {volume} {7}},\ \bibinfo {pages} {1} (\bibinfo {year} {2019})}\BibitemShut {NoStop}%
\bibitem [{\citenamefont {Aranson}(2022)}]{aranson2022bacterial}%
  \BibitemOpen
  \bibfield  {author} {\bibinfo {author} {\bibfnamefont {I.~S.}\ \bibnamefont {Aranson}},\ }\bibfield  {title} {\bibinfo {title} {Bacterial active matter},\ }\href@noop {} {\bibfield  {journal} {\bibinfo  {journal} {Reports on Progress in Physics}\ }\textbf {\bibinfo {volume} {85}},\ \bibinfo {pages} {076601} (\bibinfo {year} {2022})}\BibitemShut {NoStop}%
\bibitem [{\citenamefont {Gachelin}\ \emph {et~al.}(2014)\citenamefont {Gachelin}, \citenamefont {Rousselet}, \citenamefont {Lindner},\ and\ \citenamefont {Clement}}]{gachelin2014collective}%
  \BibitemOpen
  \bibfield  {author} {\bibinfo {author} {\bibfnamefont {J.}~\bibnamefont {Gachelin}}, \bibinfo {author} {\bibfnamefont {A.}~\bibnamefont {Rousselet}}, \bibinfo {author} {\bibfnamefont {A.}~\bibnamefont {Lindner}},\ and\ \bibinfo {author} {\bibfnamefont {E.}~\bibnamefont {Clement}},\ }\bibfield  {title} {\bibinfo {title} {Collective motion in an active suspension of escherichia coli bacteria},\ }\href@noop {} {\bibfield  {journal} {\bibinfo  {journal} {New Journal of Physics}\ }\textbf {\bibinfo {volume} {16}},\ \bibinfo {pages} {025003} (\bibinfo {year} {2014})}\BibitemShut {NoStop}%
\bibitem [{\citenamefont {Li}\ \emph {et~al.}(2021)\citenamefont {Li}, \citenamefont {Lauga},\ and\ \citenamefont {Ardekani}}]{li2021microswimming}%
  \BibitemOpen
  \bibfield  {author} {\bibinfo {author} {\bibfnamefont {G.}~\bibnamefont {Li}}, \bibinfo {author} {\bibfnamefont {E.}~\bibnamefont {Lauga}},\ and\ \bibinfo {author} {\bibfnamefont {A.~M.}\ \bibnamefont {Ardekani}},\ }\bibfield  {title} {\bibinfo {title} {Microswimming in viscoelastic fluids},\ }\href@noop {} {\bibfield  {journal} {\bibinfo  {journal} {Journal of Non-Newtonian Fluid Mechanics}\ }\textbf {\bibinfo {volume} {297}},\ \bibinfo {pages} {104655} (\bibinfo {year} {2021})}\BibitemShut {NoStop}%
\bibitem [{\citenamefont {Martinez}\ \emph {et~al.}(2014)\citenamefont {Martinez}, \citenamefont {Schwarz-Linek}, \citenamefont {Reufer}, \citenamefont {Wilson}, \citenamefont {Morozov},\ and\ \citenamefont {Poon}}]{martinez2014flagellated}%
  \BibitemOpen
  \bibfield  {author} {\bibinfo {author} {\bibfnamefont {V.~A.}\ \bibnamefont {Martinez}}, \bibinfo {author} {\bibfnamefont {J.}~\bibnamefont {Schwarz-Linek}}, \bibinfo {author} {\bibfnamefont {M.}~\bibnamefont {Reufer}}, \bibinfo {author} {\bibfnamefont {L.~G.}\ \bibnamefont {Wilson}}, \bibinfo {author} {\bibfnamefont {A.~N.}\ \bibnamefont {Morozov}},\ and\ \bibinfo {author} {\bibfnamefont {W.~C.}\ \bibnamefont {Poon}},\ }\bibfield  {title} {\bibinfo {title} {Flagellated bacterial motility in polymer solutions},\ }\href@noop {} {\bibfield  {journal} {\bibinfo  {journal} {Proceedings of the National Academy of Sciences}\ }\textbf {\bibinfo {volume} {111}},\ \bibinfo {pages} {17771} (\bibinfo {year} {2014})}\BibitemShut {NoStop}%
\bibitem [{\citenamefont {Pedersen}\ \emph {et~al.}(2013)\citenamefont {Pedersen}, \citenamefont {Vilmann}, \citenamefont {Bar-Shalom}, \citenamefont {M{\"u}llertz},\ and\ \citenamefont {Baldursdottir}}]{pedersen2013characterization}%
  \BibitemOpen
  \bibfield  {author} {\bibinfo {author} {\bibfnamefont {P.~B.}\ \bibnamefont {Pedersen}}, \bibinfo {author} {\bibfnamefont {P.}~\bibnamefont {Vilmann}}, \bibinfo {author} {\bibfnamefont {D.}~\bibnamefont {Bar-Shalom}}, \bibinfo {author} {\bibfnamefont {A.}~\bibnamefont {M{\"u}llertz}},\ and\ \bibinfo {author} {\bibfnamefont {S.}~\bibnamefont {Baldursdottir}},\ }\bibfield  {title} {\bibinfo {title} {Characterization of fasted human gastric fluid for relevant rheological parameters and gastric lipase activities},\ }\href@noop {} {\bibfield  {journal} {\bibinfo  {journal} {European Journal of Pharmaceutics and Biopharmaceutics}\ }\textbf {\bibinfo {volume} {85}},\ \bibinfo {pages} {958} (\bibinfo {year} {2013})}\BibitemShut {NoStop}%
\bibitem [{\citenamefont {Ruiz-Pulido}\ and\ \citenamefont {Medina}(2021)}]{ruiz2021overview}%
  \BibitemOpen
  \bibfield  {author} {\bibinfo {author} {\bibfnamefont {G.}~\bibnamefont {Ruiz-Pulido}}\ and\ \bibinfo {author} {\bibfnamefont {D.~I.}\ \bibnamefont {Medina}},\ }\bibfield  {title} {\bibinfo {title} {An overview of gastrointestinal mucus rheology under different ph conditions and introduction to ph-dependent rheological interactions with plga and chitosan nanoparticles},\ }\href@noop {} {\bibfield  {journal} {\bibinfo  {journal} {European Journal of Pharmaceutics and Biopharmaceutics}\ }\textbf {\bibinfo {volume} {159}},\ \bibinfo {pages} {123} (\bibinfo {year} {2021})}\BibitemShut {NoStop}%
\bibitem [{\citenamefont {Magariyama}\ and\ \citenamefont {Kudo}(2002)}]{magariyama2002mathematical}%
  \BibitemOpen
  \bibfield  {author} {\bibinfo {author} {\bibfnamefont {Y.}~\bibnamefont {Magariyama}}\ and\ \bibinfo {author} {\bibfnamefont {S.}~\bibnamefont {Kudo}},\ }\bibfield  {title} {\bibinfo {title} {A mathematical explanation of an increase in bacterial swimming speed with viscosity in linear-polymer solutions},\ }\href@noop {} {\bibfield  {journal} {\bibinfo  {journal} {Biophysical Journal}\ }\textbf {\bibinfo {volume} {83}},\ \bibinfo {pages} {733} (\bibinfo {year} {2002})}\BibitemShut {NoStop}%
\bibitem [{\citenamefont {Qu}\ and\ \citenamefont {Breuer}(2020)}]{qu2020effects}%
  \BibitemOpen
  \bibfield  {author} {\bibinfo {author} {\bibfnamefont {Z.}~\bibnamefont {Qu}}\ and\ \bibinfo {author} {\bibfnamefont {K.~S.}\ \bibnamefont {Breuer}},\ }\bibfield  {title} {\bibinfo {title} {Effects of shear-thinning viscosity and viscoelastic stresses on flagellated bacteria motility},\ }\href@noop {} {\bibfield  {journal} {\bibinfo  {journal} {Physical Review Fluids}\ }\textbf {\bibinfo {volume} {5}},\ \bibinfo {pages} {073103} (\bibinfo {year} {2020})}\BibitemShut {NoStop}%
\bibitem [{\citenamefont {Patteson}\ \emph {et~al.}(2015)\citenamefont {Patteson}, \citenamefont {Gopinath}, \citenamefont {Goulian},\ and\ \citenamefont {Arratia}}]{patteson2015running}%
  \BibitemOpen
  \bibfield  {author} {\bibinfo {author} {\bibfnamefont {A.~E.}\ \bibnamefont {Patteson}}, \bibinfo {author} {\bibfnamefont {A.}~\bibnamefont {Gopinath}}, \bibinfo {author} {\bibfnamefont {M.}~\bibnamefont {Goulian}},\ and\ \bibinfo {author} {\bibfnamefont {P.~E.}\ \bibnamefont {Arratia}},\ }\bibfield  {title} {\bibinfo {title} {Running and tumbling with e. coli in polymeric solutions},\ }\href@noop {} {\bibfield  {journal} {\bibinfo  {journal} {Scientific Reports}\ }\textbf {\bibinfo {volume} {5}},\ \bibinfo {pages} {15761} (\bibinfo {year} {2015})}\BibitemShut {NoStop}%
\bibitem [{\citenamefont {Tung}\ \emph {et~al.}(2017)\citenamefont {Tung}, \citenamefont {Lin}, \citenamefont {Harvey}, \citenamefont {Fiore}, \citenamefont {Ardon}, \citenamefont {Wu},\ and\ \citenamefont {Suarez}}]{tung2017fluid}%
  \BibitemOpen
  \bibfield  {author} {\bibinfo {author} {\bibfnamefont {C.-k.}\ \bibnamefont {Tung}}, \bibinfo {author} {\bibfnamefont {C.}~\bibnamefont {Lin}}, \bibinfo {author} {\bibfnamefont {B.}~\bibnamefont {Harvey}}, \bibinfo {author} {\bibfnamefont {A.~G.}\ \bibnamefont {Fiore}}, \bibinfo {author} {\bibfnamefont {F.}~\bibnamefont {Ardon}}, \bibinfo {author} {\bibfnamefont {M.}~\bibnamefont {Wu}},\ and\ \bibinfo {author} {\bibfnamefont {S.~S.}\ \bibnamefont {Suarez}},\ }\bibfield  {title} {\bibinfo {title} {Fluid viscoelasticity promotes collective swimming of sperm},\ }\href@noop {} {\bibfield  {journal} {\bibinfo  {journal} {Scientific Reports}\ }\textbf {\bibinfo {volume} {7}},\ \bibinfo {pages} {3152} (\bibinfo {year} {2017})}\BibitemShut {NoStop}%
\bibitem [{\citenamefont {Li}\ and\ \citenamefont {Ardekani}(2016)}]{li2016collective}%
  \BibitemOpen
  \bibfield  {author} {\bibinfo {author} {\bibfnamefont {G.}~\bibnamefont {Li}}\ and\ \bibinfo {author} {\bibfnamefont {A.~M.}\ \bibnamefont {Ardekani}},\ }\bibfield  {title} {\bibinfo {title} {Collective motion of microorganisms in a viscoelastic fluid},\ }\href@noop {} {\bibfield  {journal} {\bibinfo  {journal} {Physical Review Letters}\ }\textbf {\bibinfo {volume} {117}},\ \bibinfo {pages} {118001} (\bibinfo {year} {2016})}\BibitemShut {NoStop}%
\bibitem [{\citenamefont {Ran}\ \emph {et~al.}(2021)\citenamefont {Ran}, \citenamefont {Gagnon}, \citenamefont {Morozov},\ and\ \citenamefont {Arratia}}]{ran2021polymers}%
  \BibitemOpen
  \bibfield  {author} {\bibinfo {author} {\bibfnamefont {R.}~\bibnamefont {Ran}}, \bibinfo {author} {\bibfnamefont {D.~A.}\ \bibnamefont {Gagnon}}, \bibinfo {author} {\bibfnamefont {A.}~\bibnamefont {Morozov}},\ and\ \bibinfo {author} {\bibfnamefont {P.~E.}\ \bibnamefont {Arratia}},\ }\bibfield  {title} {\bibinfo {title} {Polymers in swarming bacterial turbulence},\ }\href@noop {} {\bibfield  {journal} {\bibinfo  {journal} {arXiv preprint arXiv:2111.00068}\ } (\bibinfo {year} {2021})}\BibitemShut {NoStop}%
\bibitem [{\citenamefont {Liao}\ and\ \citenamefont {Aranson}(2023)}]{liao2023viscoelasticity}%
  \BibitemOpen
  \bibfield  {author} {\bibinfo {author} {\bibfnamefont {W.}~\bibnamefont {Liao}}\ and\ \bibinfo {author} {\bibfnamefont {I.~S.}\ \bibnamefont {Aranson}},\ }\bibfield  {title} {\bibinfo {title} {Viscoelasticity enhances collective motion of bacteria},\ }\href@noop {} {\bibfield  {journal} {\bibinfo  {journal} {PNAS Nexus}\ }\textbf {\bibinfo {volume} {2}},\ \bibinfo {pages} {pgad291} (\bibinfo {year} {2023})}\BibitemShut {NoStop}%
\bibitem [{\citenamefont {Reinken}\ and\ \citenamefont {Menzel}(2025)}]{reinken2025self}%
  \BibitemOpen
  \bibfield  {author} {\bibinfo {author} {\bibfnamefont {H.}~\bibnamefont {Reinken}}\ and\ \bibinfo {author} {\bibfnamefont {A.~M.}\ \bibnamefont {Menzel}},\ }\bibfield  {title} {\bibinfo {title} {Self-sustained patchy turbulence in shear-thinning active fluids},\ }\href@noop {} {\bibfield  {journal} {\bibinfo  {journal} {Communications Physics}\ }\textbf {\bibinfo {volume} {8}},\ \bibinfo {pages} {270} (\bibinfo {year} {2025})}\BibitemShut {NoStop}%
\bibitem [{\citenamefont {Yang}\ \emph {et~al.}(2024)\citenamefont {Yang}, \citenamefont {Zheng}, \citenamefont {Cheng},\ and\ \citenamefont {Zhang}}]{yang2024molecular}%
  \BibitemOpen
  \bibfield  {author} {\bibinfo {author} {\bibfnamefont {G.}~\bibnamefont {Yang}}, \bibinfo {author} {\bibfnamefont {T.}~\bibnamefont {Zheng}}, \bibinfo {author} {\bibfnamefont {Q.}~\bibnamefont {Cheng}},\ and\ \bibinfo {author} {\bibfnamefont {H.}~\bibnamefont {Zhang}},\ }\bibfield  {title} {\bibinfo {title} {Molecular dynamics simulation of the flow mechanism of shear-thinning fluids in a microchannel},\ }\href@noop {} {\bibfield  {journal} {\bibinfo  {journal} {Chinese Physics B}\ }\textbf {\bibinfo {volume} {33}},\ \bibinfo {pages} {044701} (\bibinfo {year} {2024})}\BibitemShut {NoStop}%
\bibitem [{\citenamefont {Boffetta}\ and\ \citenamefont {Ecke}(2012)}]{boffetta2012two}%
  \BibitemOpen
  \bibfield  {author} {\bibinfo {author} {\bibfnamefont {G.}~\bibnamefont {Boffetta}}\ and\ \bibinfo {author} {\bibfnamefont {R.~E.}\ \bibnamefont {Ecke}},\ }\bibfield  {title} {\bibinfo {title} {Two-dimensional turbulence},\ }\href@noop {} {\bibfield  {journal} {\bibinfo  {journal} {Annual Review of Fluid Mechanics}\ }\textbf {\bibinfo {volume} {44}},\ \bibinfo {pages} {427} (\bibinfo {year} {2012})}\BibitemShut {NoStop}%
\bibitem [{\citenamefont {Kraichnan}\ and\ \citenamefont {Montgomery}(1980)}]{kraichnan1980two}%
  \BibitemOpen
  \bibfield  {author} {\bibinfo {author} {\bibfnamefont {R.~H.}\ \bibnamefont {Kraichnan}}\ and\ \bibinfo {author} {\bibfnamefont {D.}~\bibnamefont {Montgomery}},\ }\bibfield  {title} {\bibinfo {title} {Two-dimensional turbulence},\ }\href@noop {} {\bibfield  {journal} {\bibinfo  {journal} {Reports on Progress in Physics}\ }\textbf {\bibinfo {volume} {43}},\ \bibinfo {pages} {547} (\bibinfo {year} {1980})}\BibitemShut {NoStop}%
\bibitem [{\citenamefont {Cai}\ \emph {et~al.}(2018)\citenamefont {Cai}, \citenamefont {Huang}, \citenamefont {Ye},\ and\ \citenamefont {Xu}}]{cai2018dynamic}%
  \BibitemOpen
  \bibfield  {author} {\bibinfo {author} {\bibfnamefont {S.}~\bibnamefont {Cai}}, \bibinfo {author} {\bibfnamefont {Y.}~\bibnamefont {Huang}}, \bibinfo {author} {\bibfnamefont {B.}~\bibnamefont {Ye}},\ and\ \bibinfo {author} {\bibfnamefont {C.}~\bibnamefont {Xu}},\ }\bibfield  {title} {\bibinfo {title} {Dynamic illumination optical flow computing for sensing multiple mobile robots from a drone},\ }\href@noop {} {\bibfield  {journal} {\bibinfo  {journal} {IEEE Transactions on Systems, Man, and Cybernetics: Systems}\ }\textbf {\bibinfo {volume} {48}},\ \bibinfo {pages} {1370} (\bibinfo {year} {2018})}\BibitemShut {NoStop}%
\bibitem [{Bia()}]{Bian2026Supplymental}%
  \BibitemOpen
  \href@noop {} {}\bibinfo {note} {See Supplemental Material at \url{http://link.aps.org/supplemental/10.1103/z22f-3v22} for detailed experimental results and modified resistive force theory, which includes Refs.~\cite{boffetta2004lagrangian,wensink2012meso,magariyama2002mathematical,darnton2007torque,holwill1963hydrodynamic,qu2020effects,cox1970motion,kundu2024fluid}.}\BibitemShut {Stop}%
\bibitem [{\citenamefont {Dunkel}\ \emph {et~al.}(2013)\citenamefont {Dunkel}, \citenamefont {Heidenreich}, \citenamefont {Drescher}, \citenamefont {Wensink}, \citenamefont {B{\"a}r},\ and\ \citenamefont {Goldstein}}]{dunkel2013fluid}%
  \BibitemOpen
  \bibfield  {author} {\bibinfo {author} {\bibfnamefont {J.}~\bibnamefont {Dunkel}}, \bibinfo {author} {\bibfnamefont {S.}~\bibnamefont {Heidenreich}}, \bibinfo {author} {\bibfnamefont {K.}~\bibnamefont {Drescher}}, \bibinfo {author} {\bibfnamefont {H.~H.}\ \bibnamefont {Wensink}}, \bibinfo {author} {\bibfnamefont {M.}~\bibnamefont {B{\"a}r}},\ and\ \bibinfo {author} {\bibfnamefont {R.~E.}\ \bibnamefont {Goldstein}},\ }\bibfield  {title} {\bibinfo {title} {Fluid dynamics of bacterial turbulence},\ }\href@noop {} {\bibfield  {journal} {\bibinfo  {journal} {Physical Review Letters}\ }\textbf {\bibinfo {volume} {110}},\ \bibinfo {pages} {228102} (\bibinfo {year} {2013})}\BibitemShut {NoStop}%
\bibitem [{\citenamefont {Xie}\ \emph {et~al.}(2022)\citenamefont {Xie}, \citenamefont {Liu}, \citenamefont {Luo},\ and\ \citenamefont {Jing}}]{xie2022activity}%
  \BibitemOpen
  \bibfield  {author} {\bibinfo {author} {\bibfnamefont {C.}~\bibnamefont {Xie}}, \bibinfo {author} {\bibfnamefont {Y.}~\bibnamefont {Liu}}, \bibinfo {author} {\bibfnamefont {H.}~\bibnamefont {Luo}},\ and\ \bibinfo {author} {\bibfnamefont {G.}~\bibnamefont {Jing}},\ }\bibfield  {title} {\bibinfo {title} {Activity-induced enhancement of superdiffusive transport in bacterial turbulence},\ }\href@noop {} {\bibfield  {journal} {\bibinfo  {journal} {Micromachines}\ }\textbf {\bibinfo {volume} {13}},\ \bibinfo {pages} {746} (\bibinfo {year} {2022})}\BibitemShut {NoStop}%
\bibitem [{\citenamefont {Sokolov}\ and\ \citenamefont {Aranson}(2012)}]{sokolov2012physical}%
  \BibitemOpen
  \bibfield  {author} {\bibinfo {author} {\bibfnamefont {A.}~\bibnamefont {Sokolov}}\ and\ \bibinfo {author} {\bibfnamefont {I.~S.}\ \bibnamefont {Aranson}},\ }\bibfield  {title} {\bibinfo {title} {Physical properties of collective motion in suspensions of bacteria},\ }\href@noop {} {\bibfield  {journal} {\bibinfo  {journal} {Physical Review Letters}\ }\textbf {\bibinfo {volume} {109}},\ \bibinfo {pages} {248109} (\bibinfo {year} {2012})}\BibitemShut {NoStop}%
\bibitem [{\citenamefont {Wang}\ and\ \citenamefont {Huang}(2017)}]{wang2017intrinsic}%
  \BibitemOpen
  \bibfield  {author} {\bibinfo {author} {\bibfnamefont {L.}~\bibnamefont {Wang}}\ and\ \bibinfo {author} {\bibfnamefont {Y.}~\bibnamefont {Huang}},\ }\bibfield  {title} {\bibinfo {title} {Intrinsic flow structure and multifractality in two-dimensional bacterial turbulence},\ }\href@noop {} {\bibfield  {journal} {\bibinfo  {journal} {Physical Review E}\ }\textbf {\bibinfo {volume} {95}},\ \bibinfo {pages} {052215} (\bibinfo {year} {2017})}\BibitemShut {NoStop}%
\bibitem [{\citenamefont {Alert}\ \emph {et~al.}(2022)\citenamefont {Alert}, \citenamefont {Casademunt},\ and\ \citenamefont {Joanny}}]{alert2022active}%
  \BibitemOpen
  \bibfield  {author} {\bibinfo {author} {\bibfnamefont {R.}~\bibnamefont {Alert}}, \bibinfo {author} {\bibfnamefont {J.}~\bibnamefont {Casademunt}},\ and\ \bibinfo {author} {\bibfnamefont {J.-F.}\ \bibnamefont {Joanny}},\ }\bibfield  {title} {\bibinfo {title} {Active turbulence},\ }\href@noop {} {\bibfield  {journal} {\bibinfo  {journal} {Annual Review of Condensed Matter Physics}\ }\textbf {\bibinfo {volume} {13}},\ \bibinfo {pages} {143} (\bibinfo {year} {2022})}\BibitemShut {NoStop}%
\bibitem [{\citenamefont {Ishikawa}\ \emph {et~al.}(2011)\citenamefont {Ishikawa}, \citenamefont {Yoshida}, \citenamefont {Ueno}, \citenamefont {Wiedeman}, \citenamefont {Imai},\ and\ \citenamefont {Yamaguchi}}]{ishikawa2011energy}%
  \BibitemOpen
  \bibfield  {author} {\bibinfo {author} {\bibfnamefont {T.}~\bibnamefont {Ishikawa}}, \bibinfo {author} {\bibfnamefont {N.}~\bibnamefont {Yoshida}}, \bibinfo {author} {\bibfnamefont {H.}~\bibnamefont {Ueno}}, \bibinfo {author} {\bibfnamefont {M.}~\bibnamefont {Wiedeman}}, \bibinfo {author} {\bibfnamefont {Y.}~\bibnamefont {Imai}},\ and\ \bibinfo {author} {\bibfnamefont {T.}~\bibnamefont {Yamaguchi}},\ }\bibfield  {title} {\bibinfo {title} {Energy transport in a concentrated suspension of bacteria},\ }\href@noop {} {\bibfield  {journal} {\bibinfo  {journal} {Physical Review Letters}\ }\textbf {\bibinfo {volume} {107}},\ \bibinfo {pages} {028102} (\bibinfo {year} {2011})}\BibitemShut {NoStop}%
\bibitem [{\citenamefont {Qu}\ \emph {et~al.}(2018)\citenamefont {Qu}, \citenamefont {Temel}, \citenamefont {Henderikx},\ and\ \citenamefont {Breuer}}]{qu2018changes}%
  \BibitemOpen
  \bibfield  {author} {\bibinfo {author} {\bibfnamefont {Z.}~\bibnamefont {Qu}}, \bibinfo {author} {\bibfnamefont {F.~Z.}\ \bibnamefont {Temel}}, \bibinfo {author} {\bibfnamefont {R.}~\bibnamefont {Henderikx}},\ and\ \bibinfo {author} {\bibfnamefont {K.~S.}\ \bibnamefont {Breuer}},\ }\bibfield  {title} {\bibinfo {title} {Changes in the flagellar bundling time account for variations in swimming behavior of flagellated bacteria in viscous media},\ }\href@noop {} {\bibfield  {journal} {\bibinfo  {journal} {Proceedings of the National Academy of Sciences}\ }\textbf {\bibinfo {volume} {115}},\ \bibinfo {pages} {1707} (\bibinfo {year} {2018})}\BibitemShut {NoStop}%
\bibitem [{\citenamefont {Peng}\ \emph {et~al.}(2021)\citenamefont {Peng}, \citenamefont {Liu},\ and\ \citenamefont {Cheng}}]{peng2021imaging}%
  \BibitemOpen
  \bibfield  {author} {\bibinfo {author} {\bibfnamefont {Y.}~\bibnamefont {Peng}}, \bibinfo {author} {\bibfnamefont {Z.}~\bibnamefont {Liu}},\ and\ \bibinfo {author} {\bibfnamefont {X.}~\bibnamefont {Cheng}},\ }\bibfield  {title} {\bibinfo {title} {Imaging the emergence of bacterial turbulence: Phase diagram and transition kinetics},\ }\href@noop {} {\bibfield  {journal} {\bibinfo  {journal} {Science Advances}\ }\textbf {\bibinfo {volume} {7}},\ \bibinfo {pages} {eabd1240} (\bibinfo {year} {2021})}\BibitemShut {NoStop}%
\bibitem [{\citenamefont {Ershov}\ \emph {et~al.}(2022)\citenamefont {Ershov}, \citenamefont {Phan}, \citenamefont {Pylv{\"a}n{\"a}inen}, \citenamefont {Rigaud}, \citenamefont {Le~Blanc}, \citenamefont {Charles-Orszag}, \citenamefont {Conway}, \citenamefont {Laine}, \citenamefont {Roy}, \citenamefont {Bonazzi} \emph {et~al.}}]{ershov2022trackmate}%
  \BibitemOpen
  \bibfield  {author} {\bibinfo {author} {\bibfnamefont {D.}~\bibnamefont {Ershov}}, \bibinfo {author} {\bibfnamefont {M.-S.}\ \bibnamefont {Phan}}, \bibinfo {author} {\bibfnamefont {J.~W.}\ \bibnamefont {Pylv{\"a}n{\"a}inen}}, \bibinfo {author} {\bibfnamefont {S.~U.}\ \bibnamefont {Rigaud}}, \bibinfo {author} {\bibfnamefont {L.}~\bibnamefont {Le~Blanc}}, \bibinfo {author} {\bibfnamefont {A.}~\bibnamefont {Charles-Orszag}}, \bibinfo {author} {\bibfnamefont {J.~R.}\ \bibnamefont {Conway}}, \bibinfo {author} {\bibfnamefont {R.~F.}\ \bibnamefont {Laine}}, \bibinfo {author} {\bibfnamefont {N.~H.}\ \bibnamefont {Roy}}, \bibinfo {author} {\bibfnamefont {D.}~\bibnamefont {Bonazzi}}, \emph {et~al.},\ }\bibfield  {title} {\bibinfo {title} {Trackmate 7: integrating state-of-the-art segmentation algorithms into tracking pipelines},\ }\href@noop {} {\bibfield  {journal} {\bibinfo  {journal} {Nature Methods}\ }\textbf {\bibinfo {volume} {19}},\ \bibinfo {pages} {829} (\bibinfo {year} {2022})}\BibitemShut {NoStop}%
\bibitem [{\citenamefont {Boffetta}\ \emph {et~al.}(2004)\citenamefont {Boffetta}, \citenamefont {Davoudi}, \citenamefont {Eckhardt},\ and\ \citenamefont {Schumacher}}]{boffetta2004lagrangian}%
  \BibitemOpen
  \bibfield  {author} {\bibinfo {author} {\bibfnamefont {G.}~\bibnamefont {Boffetta}}, \bibinfo {author} {\bibfnamefont {J.}~\bibnamefont {Davoudi}}, \bibinfo {author} {\bibfnamefont {B.}~\bibnamefont {Eckhardt}},\ and\ \bibinfo {author} {\bibfnamefont {J.}~\bibnamefont {Schumacher}},\ }\bibfield  {title} {\bibinfo {title} {Lagrangian tracers on a surface flow: the role of time correlations},\ }\href@noop {} {\bibfield  {journal} {\bibinfo  {journal} {Physical Review Letters}\ }\textbf {\bibinfo {volume} {93}},\ \bibinfo {pages} {134501} (\bibinfo {year} {2004})}\BibitemShut {NoStop}%
\bibitem [{\citenamefont {Wensink}\ \emph {et~al.}(2012)\citenamefont {Wensink}, \citenamefont {Dunkel}, \citenamefont {Heidenreich}, \citenamefont {Drescher}, \citenamefont {Goldstein}, \citenamefont {L{\"o}wen},\ and\ \citenamefont {Yeomans}}]{wensink2012meso}%
  \BibitemOpen
  \bibfield  {author} {\bibinfo {author} {\bibfnamefont {H.~H.}\ \bibnamefont {Wensink}}, \bibinfo {author} {\bibfnamefont {J.}~\bibnamefont {Dunkel}}, \bibinfo {author} {\bibfnamefont {S.}~\bibnamefont {Heidenreich}}, \bibinfo {author} {\bibfnamefont {K.}~\bibnamefont {Drescher}}, \bibinfo {author} {\bibfnamefont {R.~E.}\ \bibnamefont {Goldstein}}, \bibinfo {author} {\bibfnamefont {H.}~\bibnamefont {L{\"o}wen}},\ and\ \bibinfo {author} {\bibfnamefont {J.~M.}\ \bibnamefont {Yeomans}},\ }\bibfield  {title} {\bibinfo {title} {Meso-scale turbulence in living fluids},\ }\href@noop {} {\bibfield  {journal} {\bibinfo  {journal} {Proceedings of the National Academy of Sciences}\ }\textbf {\bibinfo {volume} {109}},\ \bibinfo {pages} {14308} (\bibinfo {year} {2012})}\BibitemShut {NoStop}%
\bibitem [{\citenamefont {Darnton}\ \emph {et~al.}(2007)\citenamefont {Darnton}, \citenamefont {Turner}, \citenamefont {Rojevsky},\ and\ \citenamefont {Berg}}]{darnton2007torque}%
  \BibitemOpen
  \bibfield  {author} {\bibinfo {author} {\bibfnamefont {N.~C.}\ \bibnamefont {Darnton}}, \bibinfo {author} {\bibfnamefont {L.}~\bibnamefont {Turner}}, \bibinfo {author} {\bibfnamefont {S.}~\bibnamefont {Rojevsky}},\ and\ \bibinfo {author} {\bibfnamefont {H.~C.}\ \bibnamefont {Berg}},\ }\bibfield  {title} {\bibinfo {title} {On torque and tumbling in swimming escherichia coli},\ }\href@noop {} {\bibfield  {journal} {\bibinfo  {journal} {Journal of Bacteriology}\ }\textbf {\bibinfo {volume} {189}},\ \bibinfo {pages} {1756} (\bibinfo {year} {2007})}\BibitemShut {NoStop}%
\bibitem [{\citenamefont {Holwill}\ and\ \citenamefont {Burge}(1963)}]{holwill1963hydrodynamic}%
  \BibitemOpen
  \bibfield  {author} {\bibinfo {author} {\bibfnamefont {M.}~\bibnamefont {Holwill}}\ and\ \bibinfo {author} {\bibfnamefont {R.}~\bibnamefont {Burge}},\ }\bibfield  {title} {\bibinfo {title} {A hydrodynamic study of the motility of flagellated bacteria},\ }\href@noop {} {\bibfield  {journal} {\bibinfo  {journal} {Archives of Biochemistry and Biophysics}\ }\textbf {\bibinfo {volume} {101}},\ \bibinfo {pages} {249} (\bibinfo {year} {1963})}\BibitemShut {NoStop}%
\bibitem [{\citenamefont {Cox}(1970)}]{cox1970motion}%
  \BibitemOpen
  \bibfield  {author} {\bibinfo {author} {\bibfnamefont {R.~G.}\ \bibnamefont {Cox}},\ }\bibfield  {title} {\bibinfo {title} {The motion of long slender bodies in a viscous fluid part 1. general theory},\ }\href@noop {} {\bibfield  {journal} {\bibinfo  {journal} {Journal of Fluid mechanics}\ }\textbf {\bibinfo {volume} {44}},\ \bibinfo {pages} {791} (\bibinfo {year} {1970})}\BibitemShut {NoStop}%
\bibitem [{\citenamefont {Kundu}\ \emph {et~al.}(2024)\citenamefont {Kundu}, \citenamefont {Cohen}, \citenamefont {Dowling},\ and\ \citenamefont {Capecelatro}}]{kundu2024fluid}%
  \BibitemOpen
  \bibfield  {author} {\bibinfo {author} {\bibfnamefont {P.~K.}\ \bibnamefont {Kundu}}, \bibinfo {author} {\bibfnamefont {I.~M.}\ \bibnamefont {Cohen}}, \bibinfo {author} {\bibfnamefont {D.~R.}\ \bibnamefont {Dowling}},\ and\ \bibinfo {author} {\bibfnamefont {J.}~\bibnamefont {Capecelatro}},\ }\href@noop {} {\emph {\bibinfo {title} {Fluid mechanics}}}\ (\bibinfo  {publisher} {Elsevier},\ \bibinfo {year} {2024})\BibitemShut {NoStop}%
\end{thebibliography}
\end{document}

\title{Supplementary Information for: Active Turbulence in Shear Thinning Fluid}

\maketitle

\section{Dilute Bacteria Velocity}
\begin{figure}[h]
    \centering
    \includegraphics[width=0.8\linewidth]{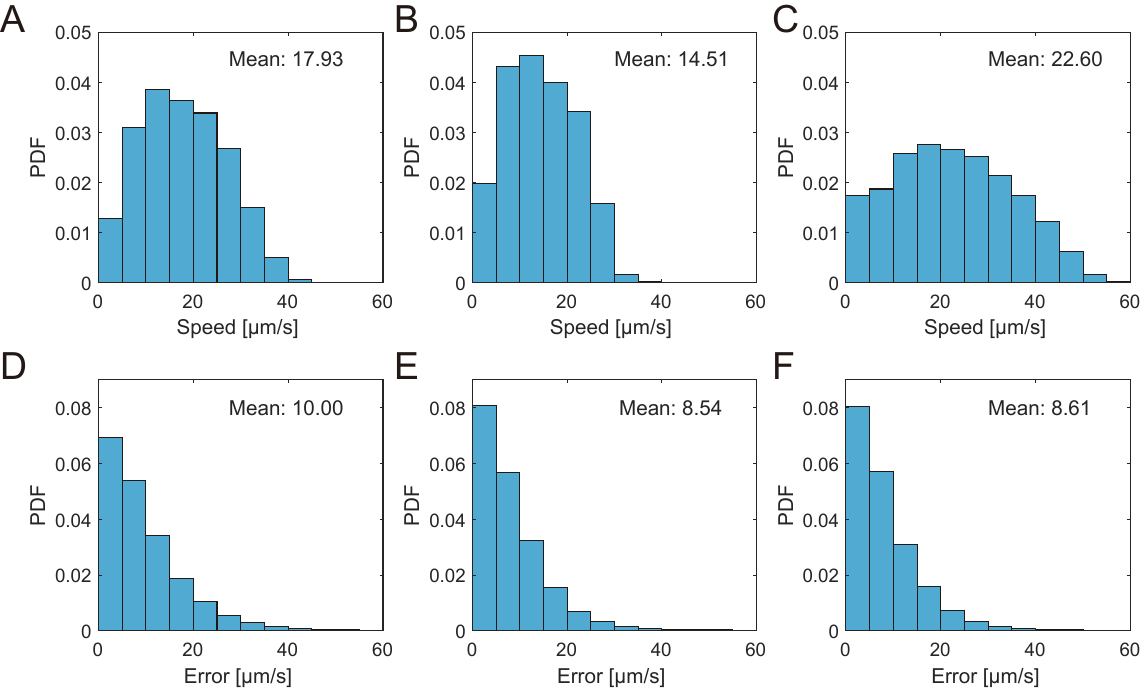}
    \caption{Velocity distribution and velocity difference of the dilute bacteria suspension. A) Smoothed speed distribution of bacteria in Buffer corresponding with the error in D). Error over mean speed is 0.56. B) Smoothed speed distribution of bacteria in 7.5\% Ficoll solution corresponding with the error in E). Error over mean speed is 0.58. C) Smoothed speed distribution of bacteria in 0.25\% Methocel solution corresponding with the error in F). Error over mean speed is 0.38. }
    \label{dilutespeed}
\end{figure}

Here we report the speed distribution of the bacteria in the dilute suspension. While the tracking is based on ImageJ, the error is introduced into the trajectory by the wobbling of the bacteria because ImageJ usually tracks the head or tail of the bacteria if the bacteria is long. Thus, the Gauss filter is applied to smoothen the original data. Here, the data with a time duration less than 1 second is excluded, and the size of the Gaussian kernel is set to 20. The speed is calculated by averaging the instantaneous speed while for the main manuscript, the speed is averaged from the repeated measurements. The distribution of the smoothed speed is shown in the Fig.\ref{dilutespeed} A, B and C. Also, the speed error (difference between the smoothed speed and the speed directly calculated from original data) can be used as an index to represent the influence of the bacterial wobble angle. The error is normalized by the mean speed and the influence of the wobble of bacteria in the Methocel solution is the smallest (0.38) compared with bacteria in buffer and Ficoll solution (0.56 and 0.58) which shows the similar results because the Newtonian fluid does not influence wobbling much. These results support the observation of the smoother path in the Methocel solution.

Experiments was conducted to compare bacterial activity with and without adding Ficoll to the bacterial suspension after re-dilution. The results show that the bacterial swimming speed in pure buffer is much lower than the speed in 1\% Ficoll solution, which shows a clear recovery of bacterial activity.

\begin{table}[h]
\centering
\caption{The speed of the bacteria in re-diluted suspension.}
\label{redilutespeed}
\begin{tabular}{lll}
\hline
\hline
Solution & Averaged bacteria swimming speed ($\mu m/s$) & Standard derivation ($\mu m/s$)\\
\hline
Buffer & 7.49531 & 2.45172\\
1\% Ficoll solution & 10.91223 & 3.92715\\
\hline
\hline
\end{tabular}
\end{table}

\section{Filter Size Selection}

\begin{figure}[h]
    \centering
    \includegraphics[width=0.5\linewidth]{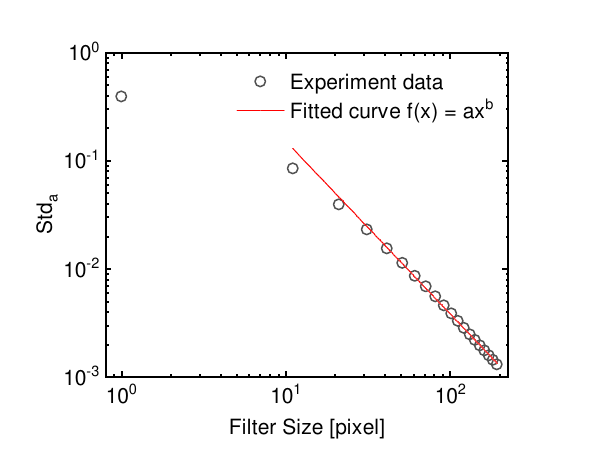}
    \caption{The Standard deviation of acceleration averaged from 50 frames of the Euler velocity field varies with the width of the Gaussian kernel. The frames are from the experiments in pure buffer. }
    \label{STDa}
\end{figure}

\begin{figure}[h]
    \centering
    \includegraphics[width=0.85\linewidth]{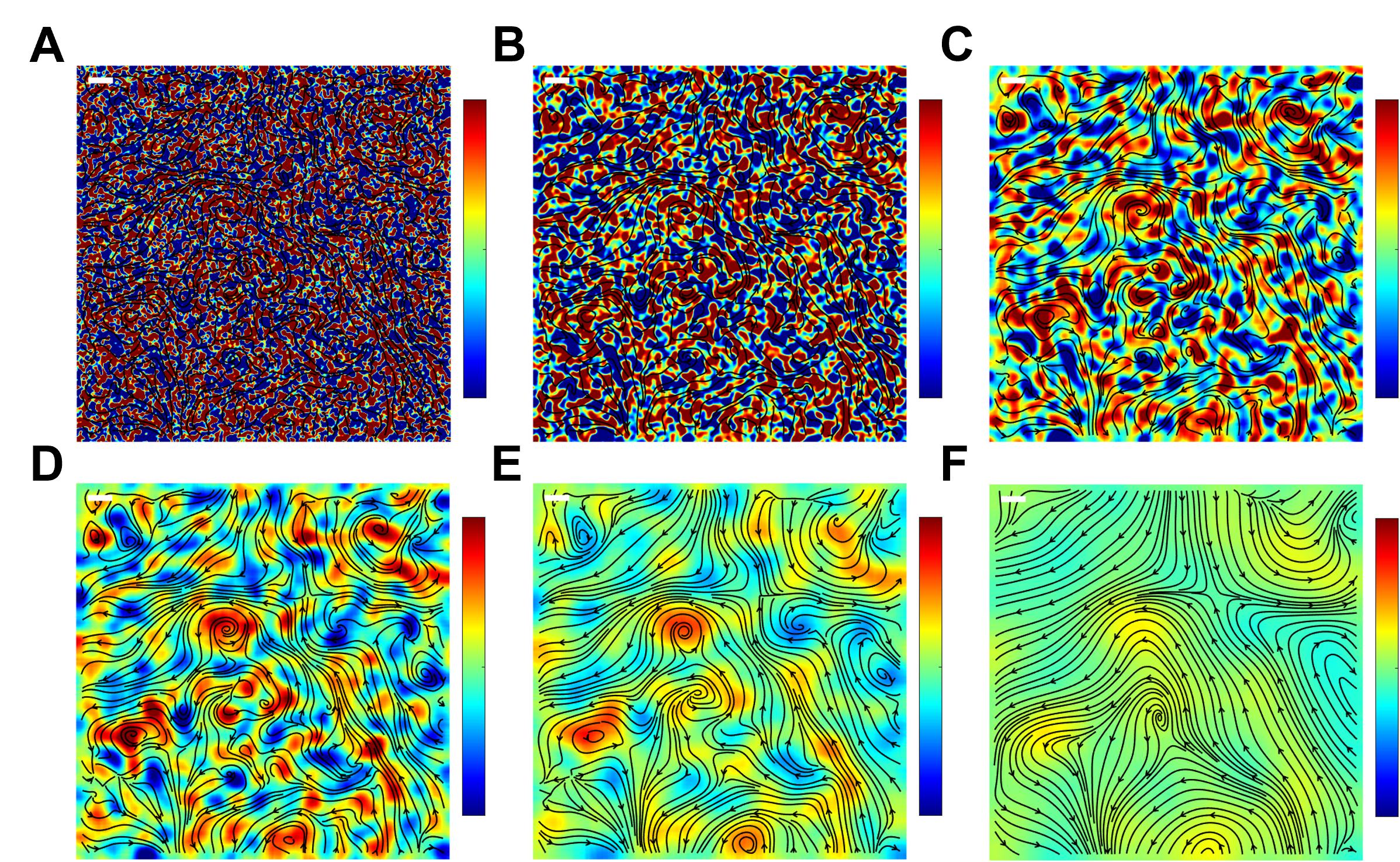}
    \caption{The vorticity field and streamline of a certain frame varies with the width of the Gaussian kernel. The filter size is A) 5, B) 10, C) 20, D) 30, E) 50, F) 100. The colorbars all denote the vorticity ranging from $-3~\mathrm{s^{-1}}$ to $3~\mathrm{s^{-1}}$, and the scale bar (white block) is $10 ~\mathrm{\mu m}$.}
    \label{VorticityField}
\end{figure}

We use the optical flow method to capture the flow velocity field of the bacteria turbulence. Then, if we directly use the raw data, the error would influence the results a lot when we calculate the spatial gradient of the velocity. So, we firstly use the Gauss filter to preprocess the data. To eliminate the noise and retain the useful data, we calculate the standard derivation of the acceleration according to the different filter size. Here, the acceleration is $\frac{Du}{Dt} = \frac{\partial u}{\partial t} + u\cdot\nabla u$. For convenience, we calculate $a' = u\cdot\nabla u$. Then, the standard derivation is $\sigma_a = \sqrt{\text{Var}(a_x)+\text{Var}(a_y)}$. When the filtering radius is small, the process primarily removes measurement noise (assumed to be white noise), causing the standard deviation of acceleration to decay exponentially. As the filtering scale increases, it gradually smooths the underlying signal itself, resulting in power-law decay. From Fig.\ref{STDa}, we can see that when filter size is less than 30, the filtering removes most of the noise. However, when it is greater than 30, the standard deviation follows a power-law decay, indicating that filtering has removed most of the necessary information. Therefore, the optimal filter size is 30. Also, Fig.\ref{VorticityField} shows the vorticity field under different selection of the filter size. Figure\ref{VorticityField} D gives the best result. 

\section{Compressibility Ratio}

\begin{figure}[h]
    \centering
    \includegraphics[width=0.5\linewidth]{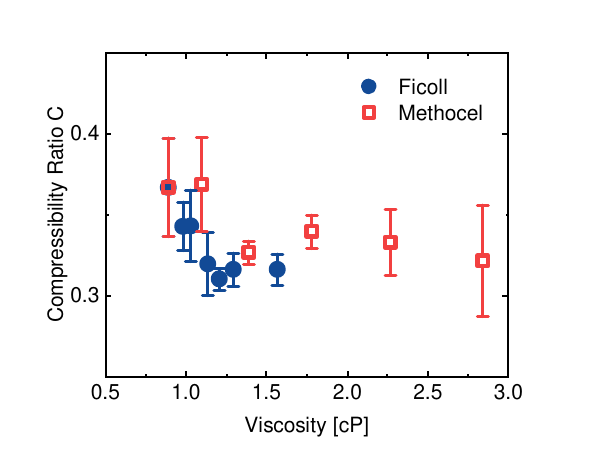}
    \caption{The compressibility ratio}
    \label{CR}
\end{figure}

Here, we calculate the compressibility ratio of the 2D velocity field to verify the 3D bacteria turbulence. The compressibility ratio $C$ is expressed as \cite{boffetta2004lagrangian} 
$$C = \frac{ \langle{ \left( \partial_x u_x + \partial_y u_y \right)^2 }\rangle }{ \langle{ \left( \partial_x u_x \right)^2 + \left( \partial_y u_x \right)^2 + \left( \partial_x u_y \right)^2 + \left( \partial_y u_y \right)^2 }\rangle }
$$
and shown in Fig.\ref{CR}. The values ranges from $0.3$ to $0.4$ and indicates the structure of 3D. 

\section{Enstrophy and Shear Rate}

\begin{figure}[h]
  \begin{minipage}[t]{0.5\linewidth}
    \centering
    \includegraphics[width=0.9\linewidth]{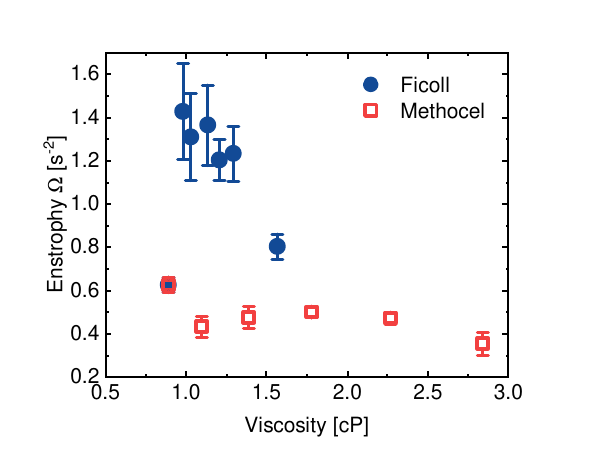}
    \caption{Enstrophy.}
    \label{ENSTRO_FIG}
  \end{minipage}%
  \begin{minipage}[t]{0.5\linewidth}
    \centering
    \includegraphics[width=0.9\linewidth]{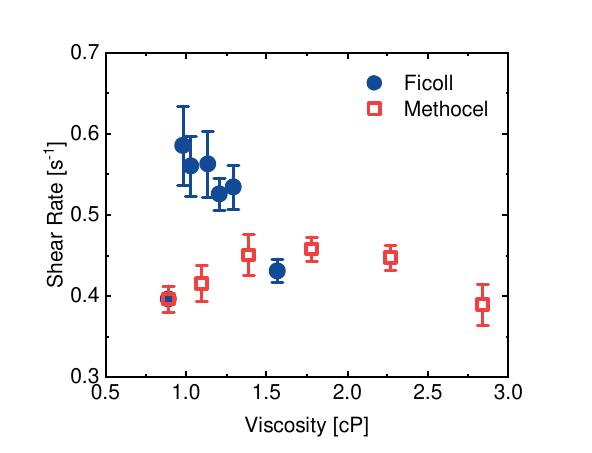}
    \caption{Shear rate.}
    \label{MSR_FIG}
  \end{minipage}
\end{figure}

The in-plane enstrophy calculated by $\Omega = \langle \omega^2/2 \rangle$, which represents the strength of rotation, is shown in Fig.\ref{ENSTRO_FIG}. The variation of enstrophy with polymer concentration is highly similar to the shear rate except for the buffer solution at viscosity of $0.89~cP$.

\section{Correlation Function}
\begin{figure}[h]
    \centering
    \includegraphics[width=0.73\linewidth]{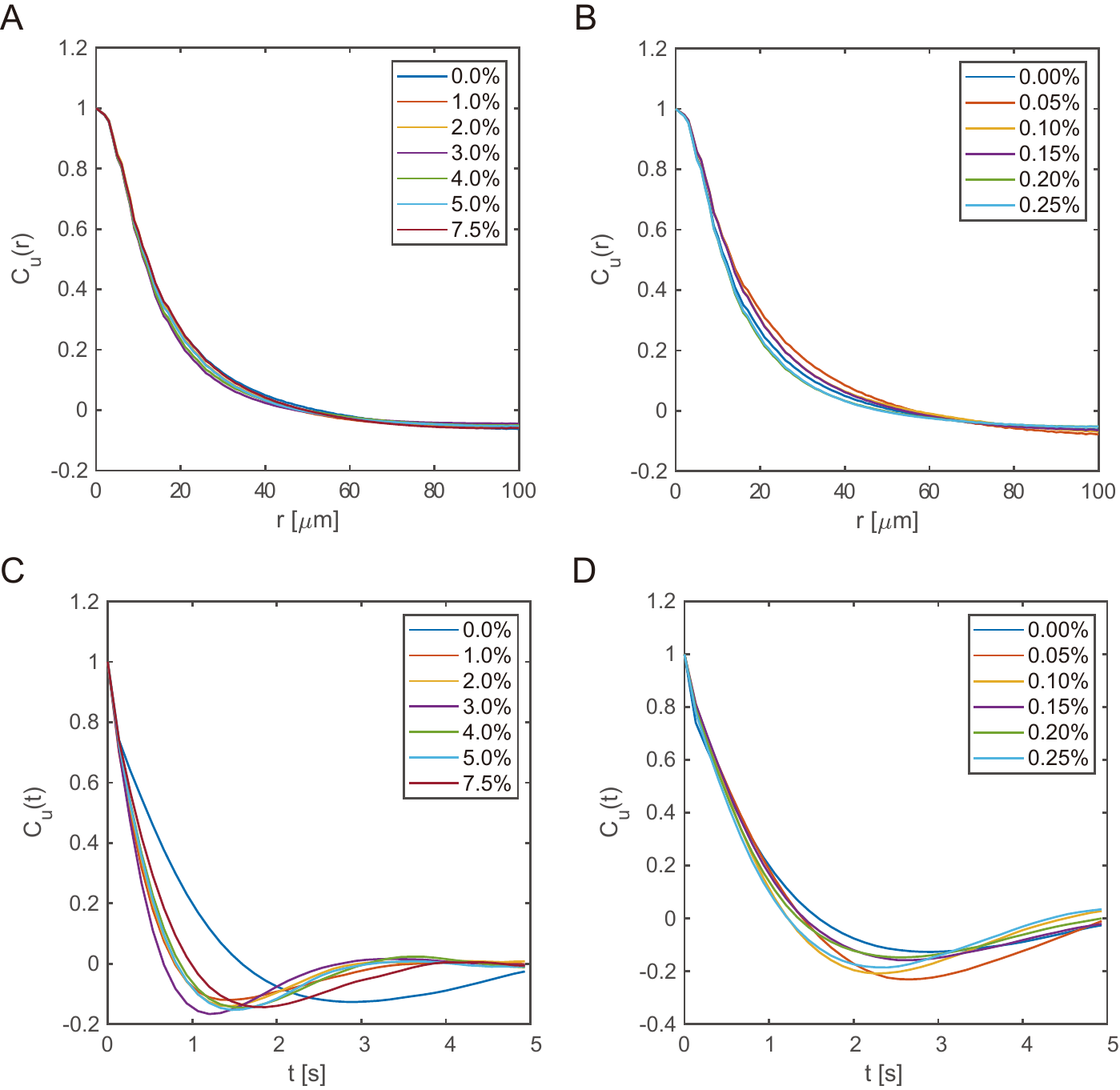}
    \caption{The spatial correlation function of velocity in A) Ficoll solution and B) Methocel solution. The temporal correlation function of velocity in C) Ficoll solution and D) Methocel solution.}
    \label{CF}
\end{figure}

\begin{figure}[h]
    \centering
    \includegraphics[width=0.73\linewidth]{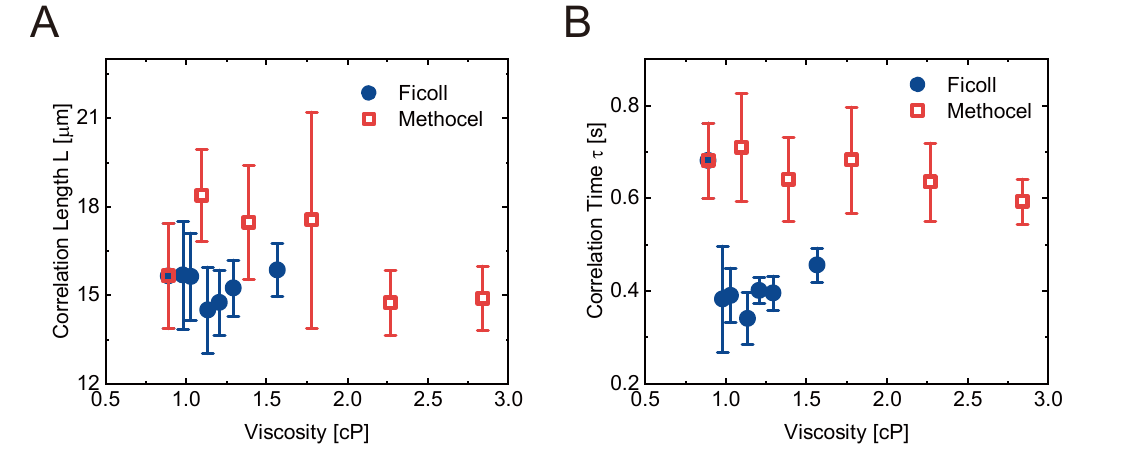}
    \caption{The velocity correlation A) length and B) time.}
    \label{CV}
\end{figure}

As dedfined in the main text, the spatial correlation function of velocity is defined as $C_u(r) = \langle \hat{v}(r_0)\cdot \hat{v}(r_0+r) \rangle_{r_0}$ and the temporal correlation function is $C_u(t) = \langle \hat{v}(t_0)\cdot \hat{v}(t_0+t) \rangle_{t_0}$. In Fig.\ref{CF}, the correlation functions decreases to the value lower than 0 and then turns to be constant around 0. The temporal correlation shows the lowest value around -0.2 while the spatial one more close to 0, which may indicate that there exists some damping of the bacteria turbulence along time scale. The corresponding correlation length and time (Fig.\ref{CV}) are grabbed at $C_u = 1/e$.

\section{Shear Viscosity of Polymer Solutions}
\begin{figure}[h]
    \centering
    \includegraphics[width=0.75\linewidth]{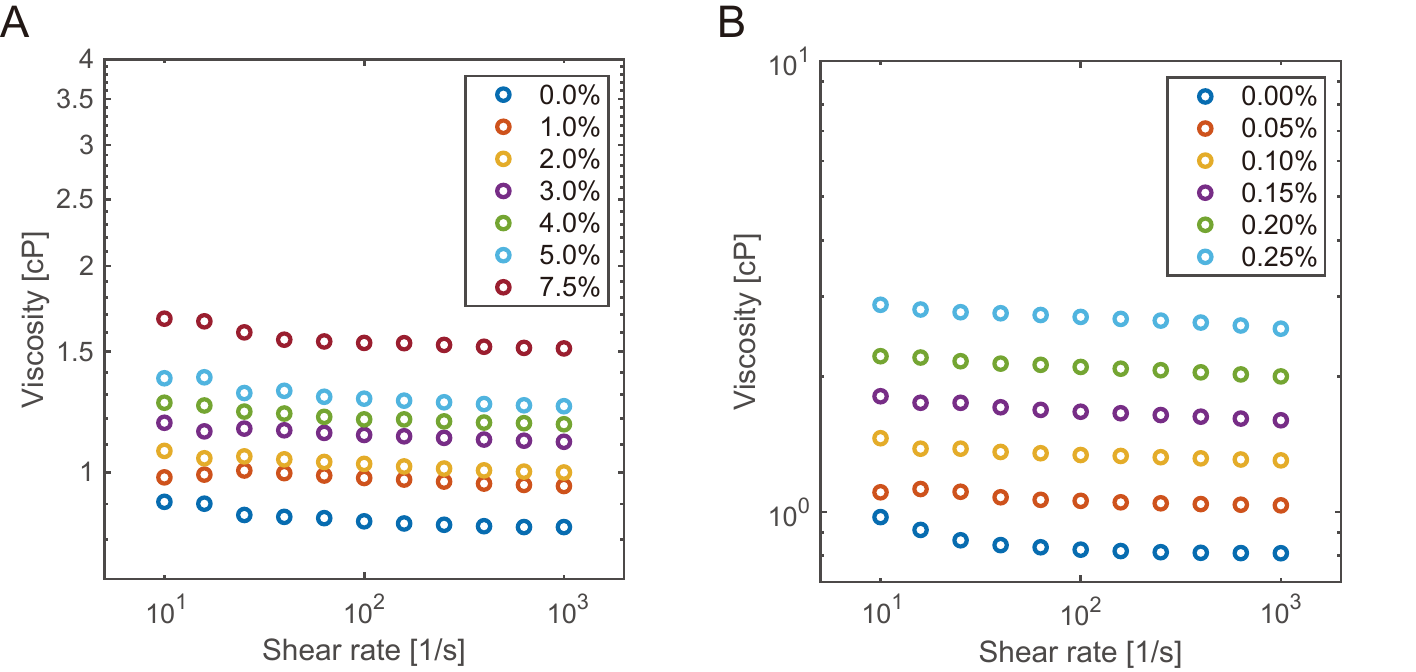}
    \caption{The shear viscosity of bacteria turbulence in polymer solutions. A) Ficoll solution. B) Methocel solution.}
    \label{PV}
\end{figure}

The shear viscosity of the Ficoll and Methocel solution without bacteria (Fig.\ref{PV}) is measured by the TA Discovery HR20 multi-purpose rheometer using a 40.0 mm $1.0^\circ$ cone plate. In our testing range, although Methocel is generally considered a shear-thinning fluid, variation of viscosity is not significant. The differences observed in the bacterial experiments should be caused by shear-thinning effects in microscopic scale.

\section{Energy Spectrum}

\begin{figure}[h]
    \centering
    \includegraphics[width=0.95\linewidth]{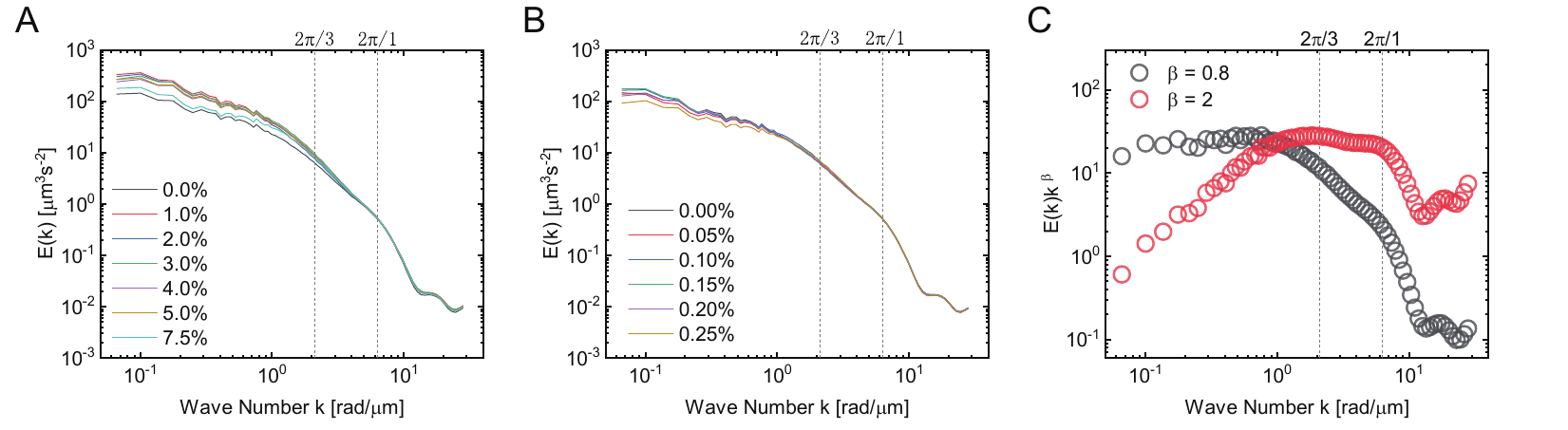}
    \caption{The energy spectrum in different polymer solutions. The two dashed lines represent the wave number for $1~\mathrm{\mu m}$ and $3~\mathrm{\mu m}$ which are the diameter and length of the bacteria. A) Ficoll solution. B) Methocel solution. C) The energy spectrum of the bacteria turbulence in pure buffer times $k^\beta$ to find the slope in log-log plot.}
    \label{EnergySpectrum}
\end{figure}

The Energy Spectrum can be obtained from the Fourier transformation of the equal-time two-point velocity correlation function under different dimension $d$ \cite{wensink2012meso}: 
$$E_d(k) = \frac{k^{d-1}}{C_d}\int d^d Re^{-i\mathbf{k}\cdot\mathbf{R}} \langle \mathbf{v}(t,\boldsymbol{r}) \cdot \mathbf{v}(t,\boldsymbol{r}+\boldsymbol{R}) \rangle,$$
where $C_d = 2 \pi$ for 2D case and $C_d = 4 \pi$ for 3D case. Here, $\langle \mathbf{v}(t,\boldsymbol{r}) \cdot \mathbf{v}(t,\boldsymbol{r}+\boldsymbol{R}) \rangle$ is the normalized velocity correlation functions which can be used to grab the correlation length. From Fig.\ref{EnergySpectrum} A and B, we can find the difference of the energy spectrum mainly exists at the small wave number. For the large wave number, the noise might dominate the small size thus the lines overlap with each other since the analysis is from the raw data for the energy spectrum. Then, we plot $E(k)k^\beta$ for the turbulence in pure buffer and find that the slope at small wave number is around $-0.8$ while at large wave number it is around $-2$.

\section{Modified Resistive Force Theory}

\begin{table}[h]
\centering
\caption{Parameters used in bacterial swimming speed calculation. Values are obtained from \cite{magariyama2002mathematical,darnton2007torque}}
\label{tab:parameters}
\begin{tabular}{lll}
\hline
\hline
\multicolumn{1}{c}{\textbf{Symbol}} & \multicolumn{1}{c}{\textbf{Description}} & \multicolumn{1}{c}{\textbf{Value}} \\ \hline
$F_c, F_f$ & Force acting on the cell body and the flagella by fluid &  \\
$T_c, T_f$ & Torque acting on the cell body and the flagella by fluid &  \\
$T_m$ & Torque generated by the flagellar motor &  \\
$\omega_c, \omega_f, \omega_m$ & Rotation rate of the cell body, flagella and flagellar motor &  \\
$\alpha_c, \beta_c,\alpha_f, \beta_f, \gamma_f$ & Drag coefficients of cell body and flagella  &  \\
$\mu_T, \mu_N$ & viscosity along tangential and normal direction &  \\
$\tau_r$ & characteristic relaxation time for the Methocel solution &  \\
$\mu_{T,f}, \mu_{N,f}$ & effective viscosity around flagella consider viscoelasticity&  \\
$\mu_0$ & viscosity of the solution without polymer & 0.89 cP \\
$a$ & Half of the cell body width & 0.3 $\mu$m \\
$b$ & Half of the cell body length & 1 $\mu$m \\
$d$ & Radius of the flagellar filament & 0.015 $\mu$m \\
$L$ & Length of the flagellar filament & 8 $\mu$m \\
$p$ & Pitch of the flagellar helix & 2 $\mu$m \\
$r$ & Radius of the flagellar helix & 0.14 $\mu$m \\
$T_0$ & Maximal flagellar motor torque & 1.5 pN$\cdot\mu$m \\
$\omega_1$ & Rotational transition speed of flagellar motor& 90$\times 2\pi$ rad/s \\
$\omega_2$ & Maximal flagellar motor rotation speed & 180$\times 2\pi$ rad/s \\
% $\mu_0$ & Solvent viscosity (water) & 0.00089 Pa$\cdot$s \\
% $m(c)$ & Consistency index (power-law fluid) & $2.257 \times 10^5 c^2 + 299 c + 0.9081$ mPa$\cdot$s$^{n}$ \\
% $n(c)$ & Flow behavior index (power-law fluid) & $-13.18 c + 1$ (dimensionless) \\
% $\text{shearbody}$ & Shear rate at the cell body & 200 s$^{-1}$ \\
% $\text{shearflagellar}$ & Shear rate at the flagellum & 10000 s$^{-1}$ \\
$\zeta$ & Shear-thinning effect parameter & 0.4, 0.8, 1 (Newtonian) \\ 
\hline
\hline
\end{tabular}
\end{table}

This approach follows Magariyama \textit{et al.} \cite{magariyama2002mathematical}. The parameters are summarized in Table \ref{tab:parameters}. First, we show the equation of motion:
$$F_c + F_f = 0,$$ $$T_c - T_m = 0,$$ $$T_f+T_m = 0.$$

Consider the drag force and torque acting on the cell body by fluid:
$$F_c = \alpha_c v,$$ $$T_c = \beta_c \omega_c.$$

Drag force and torque acting on the flagella by fluid:
$$F_f = \alpha_f v + \gamma_f \omega_f,$$ $$T_f = \gamma_f v + \beta_f \omega_f,$$

where the drag coefficients are calculated by modifying the traditional Resistive Force Theory (RFT)  \cite{holwill1963hydrodynamic}
$$\alpha_c = -6\pi \mu_{N} a \left(1 - \frac{1}{5}\left(1 - \frac{b}{a}\right)\right)$$
$$\beta_c = -8\pi \mu_{T} a^3 \left(1 - \frac{3}{5}\left(1 - \frac{b}{a}\right)\right)$$
$$\alpha_f = \frac{2\pi \mu_{N} L}{\left(\ln\left({d}/{2p}\right) + {1}/{2}\right)\left(4\pi^2 r^2 + p^2\right)} \left(8\pi^2 r^2 + \frac{\mu_{T}}{\mu_{N}} p^2\right)$$
$$\beta_f = \frac{2\pi \mu_{N} L}{\left(\ln\left({d}/{2p}\right) + {1}/{2}\right)\left(4\pi^2 r^2 + p^2\right)} \left(2p^2 + \frac{\mu_{T}}{\mu_{N}} 4\pi^2 r^2\right) r^2$$
$$\gamma_f = \frac{2\pi \mu_{N} L}{\left(\ln\left({d}/{2p}\right) + {1}/{2}\right)\left(4\pi^2 r^2 + p^2\right)} \left(2 - \frac{\mu_{T}}{\mu_{N}}\right) \left(-2\pi r^2 p\right)$$

Different from the RFT, Magariyama \textit{et al.} \cite{magariyama2002mathematical} consider the viscosity to be anisotropic along the tangential and normal direction. Here, we simply relate the viscosity along two direction by $\mu_T = \mu_0 + \zeta(\mu_N-\mu_0)$, where $\mu_N$ is the viscosity measured by macroscopic rheological experiments. $\zeta$ describes the strength of shear-thinning.

Then, the equation of motion is rewritten as 
\begin{equation}
    \centering\label{eq1}
    \left[\begin{array}{ccc} 
        \alpha_c+\alpha_f &    0    & \gamma_f \\ 
        \gamma_f &    \beta_c   & \beta_f
    \end{array}\right]
    \left[\begin{array}{c} 
        v\\ 
        \omega_c\\ 
        \omega_f
    \end{array}\right]
    =
    \left[\begin{array}{c} 
        0 \\ 
        0
    \end{array}\right],
\end{equation}

While there are three unknown parameters, one more equation is required. Refer to the previous measurement on the flagellar motor, two regions are defined on the torque generated: constant torque region at low rotation rate and monotonically decreasing torque region with rotation rate. That gives:

$$
T_m = 
\begin{cases}
T_0 & \text{for } \omega_m \leq \omega_1 \\
(\omega_m-\omega_2)T_0/(\omega_1-\omega_2) & \text{for } \omega_1 < \omega_m \leq \omega_2
\end{cases},
$$
where $\omega_m = \omega_f-\omega_c$. Combining $T_c - T_m = 0$ with Eq.\ref{eq1} gives the complete equation set to solve the problem. 

The normal stress applied to the bacteria flagella while rotating due to viscoelasticity \cite{qu2020effects} is also considered to contribute to the effective normal viscosity denoted as $\mu_{N,f}$. Here we consider a simple case: regard the flagella as a solid cylinder with length $L$ and radius $r$ rotating at rotation rate $\omega_f = 100\times 2\pi~\mathrm{rad/s}$ and move along the axial direction of the cylinder at $v = 20 ~\mathrm{\mu m/s}$. The shear-induced normal stress \cite{qu2020effects} can be calculated by 

$$f_n = 2 \tau_r (\mu_N - \mu_0)\omega_f^2.$$ 

Then, apply the slender-body theory \cite{cox1970motion} and extreme case ($R_2 = \infty$) for the steady flow between concentric rotating cylinders \cite{kundu2024fluid}, the viscous stress can be expressed by

$$\sigma_{max} = \sqrt{(\frac{\mu_Nv}{r\text{ln}(L/r)})^2+(2 \mu_N w_f)^2}$$

The ratio between the shear-induced normal stress and viscous stress can be estimated from this simple case. Finally, we apply the modified viscosity around the flagella

$$\mu_{N,f} = \mu_N(1+\frac{f_n}{\sigma_{max}})$$
$$\mu_{T,f} = \mu_0 + \zeta(\mu_{N,f}-\mu_0)$$

to the general viscosity in coefficients $\alpha_f, \beta_f, \gamma_f$. We do not change the viscosity around the cell body since shear-induced normal stress generated by rotation of cell body is much lower than the viscous stress.

\bibliography{references}